\documentclass[twocolumn,12pt]{autart}

\usepackage[dvips]{epsfig}
\usepackage{amsmath}
\usepackage{ntheorem}
\usepackage{amssymb}
\usepackage{mathtools}
\usepackage{amsfonts}
\usepackage{color}
\usepackage{subfigure}
\usepackage{algorithm}
\usepackage{uri}
\usepackage{textcomp}
\usepackage{siunitx}
\usepackage{wrapfig}
\usepackage{hyperref}

\definecolor{gray}{rgb}{0.5, 0.5, 0.5}
\definecolor{mygreen}{RGB}{0, 204, 0}
\definecolor{urlblue}{RGB}{0, 0, 238}
\definecolor{aoenglish}{rgb}{0.0, 0.5, 0.0}
\definecolor{darkblue}{rgb}{0.0, 0.0, 0.55}
\definecolor{darkmagenta}{rgb}{0.55, 0.0, 0.55}
\definecolor{electricviolet}{rgb}{0.56, 0.0, 1.0}
\definecolor{electricyellow}{rgb}{1.0, 1.0, 0.0}
\definecolor{forestgreen}{rgb}{0.13, 0.55, 0.13}
\definecolor{fuchsia}{rgb}{1.0, 0.0, 1.0}
\definecolor{gamboge}{rgb}{0.89, 0.61, 0.06}
\definecolor{goldenpoppy}{rgb}{0.99, 0.76, 0.0}
\definecolor{indigo}{rgb}{0.29, 0.0, 0.51}
\definecolor{internationalorange}{rgb}{1.0, 0.31, 0.0}
\definecolor{lava}{rgb}{0.81, 0.06, 0.13}
\definecolor{selectiveyellow}{rgb}{1.0, 0.73, 0.0}
\definecolor{turquoiseblue}{RGB}{0,191,255}
\definecolor{darkgreen}{RGB}{0, 100, 0}

\hypersetup{
    colorlinks=true,
    linkcolor=red,
    filecolor=magenta,     
    citecolor=mygreen,
    urlcolor=urlblue,
    pdfpagemode=FullScreen,
    }

\newenvironment{proof}{\paragraph*{Proof.}}{\hfill$\square$}
\newenvironment{proofnp}{\textbf{Proof }}{\hfill$\square$}
\newtheorem{assumption}{Assumption}
\newtheorem{lemma}{Lemma}
\newtheorem{theorem}{Theorem}
\newtheorem{proposition}{Proposition}
\newtheorem{definition}{Definition}
\newtheorem{problem}{Problem}
\newtheorem{remark}{Remark}

\usepackage{enumitem}
\renewcommand{\theenumi}{\arabic{enumi}}

\renewcommand{\theenumii}{\arabic{enumii}}

\renewcommand{\theenumiii}{\arabic{enumiii}}

\newcommand{\aseq}{~\displaystyle{\mathop{=}^{\cdot}}~}

\newcommand{\E}{{\mathbb E}}
\newcommand{\R}{{\mathbb R}}

\begin{document}

\begin{frontmatter}

\title{Harnessing Uncertainty for a Separation Principle in \\ Direct Data-Driven Predictive Control\thanksref{footnoteinfo}}

\thanks[footnoteinfo]{This project was partially supported by the Italian Ministry of University and Research under the PRIN’17 project \textquotedblleft Data-driven learning of constrained control systems\textquotedblright, contract no. 2017J89ARP. Corresponding author: Alessandro Chiuso (e-mail: alessandro.chiuso@unipd.it).}

\author[UNIPD]{Alessandro Chiuso},
\author[UNIPD]{Marco Fabris},
\author[TUe]{Valentina Breschi},
\author[POLIMI]{Simone Formentin}

\address[UNIPD]{Department of Information Engineering, University of Padova, Via Gradenigo 6/b, 35131 Padova, Italy.}
\address[TUe]{Department of Electrical Engineering, Eindhoven University of Technology, 5600 MB Eindhoven, The Netherlands.}
\address[POLIMI]{Dipartimento di Elettronica, Informazione e Bioingegneria, Politecnico di Milano, P.za L. Da Vinci, 32, 20133 Milano, Italy.}




\begin{abstract}
 Model Predictive Control (MPC) is a powerful method for complex system regulation, but its reliance on an accurate model poses many limitations in real-world applications. Data-driven predictive control (DDPC) aims at overcoming this limitation, by relying on historical data to provide information on the plant to be controlled. 
In this work, we present a unified stochastic framework for direct DDPC, where control actions are obtained by optimizing the Final Control Error (FCE), which is directly computed from available data only and automatically weighs the impact of uncertainty on the control objective. 
Our framework allows us to establish a separation principle for Predictive Control, elucidating the role that predictive models and their uncertainty play in DDPC. Moreover, it generalizes existing DDPC methods, like regularized Data-enabled Predictive Control (DeePC) and $\gamma$-DDPC, providing a path toward noise-tolerant data-based control with rigorous optimality guarantees. The theoretical investigation is complemented by a series of experiments (\href{https://github.com/marcofabris92/a-separation-principle-in-d3pc}{code available on GitHub}), revealing that the proposed method consistently outperforms or, at worst, matches existing techniques without requiring tuning regularization parameters as other methods do. 



\end{abstract}

\end{frontmatter}

\section{Introduction}\label{Sec:intro}

Model Predictive Control (MPC) has earned recognition as a powerful technology for optimizing the regulation of complex systems, owing to its flexible formulation and constraint-handling capabilities \cite{morari1999model}. However, its effectiveness is contingent on the accuracy of the predictor based on which control actions are optimized \cite{borrelli2017predictive}. This limitation has led to the exploration of robust, stochastic, and tube-based MPC solutions \cite{rakovic2016model}. Unfortunately, these extensions often come with trade-offs, such as conservatism in control and substantial computational burdens, rendering them less suitable for real-time applications like mechatronics or automotive systems \cite{shi2021advanced}.\\
In response to these challenges, data-driven predictive control (DDPC), sometimes referred to as Data-enabled Predictive Control (DeePC), has emerged as an alternative to traditional MPC, see \cite{berberich2020data,breschi2022role,coulson2019data}. {DDPC directly maps data collected offline onto the control sequence starting from the current measurements, without the need for an intermediate identification phase. In the linear time-invariant setting, mathematical tools such as the ``fundamental lemma'' \cite{willems2005note} and linear algebra-based subspace and projection methods \cite{Vanov-book} represent the enabling technology for data-driven control \cite{breschi2022role,dorfler2022bridging} also providing the link between DDPC and Subspace Predictive Control \cite{FAVOREEL1999} and, more in general, between \textquotedblleft indirect\textquotedblright \ and \textquotedblleft direct\textquotedblright, \textquotedblleft model-based\textquotedblright \ and \textquotedblleft model-free\textquotedblright \ data-driven predictive control schemes \cite{DorflerIEEECSM2023}. In turn, unveiling this link has led to quite a bit of debate in the recent literature regarding the pros and cons of exploiting models (explicitly or implicitly) for control design (see e.g., the recent works \cite{dorfler2022bridging,DorflerIEEECSM2023,krishnan2021direct}), a topic that closely relates to past work on experiment design \cite{Hjalmasson-05}.\\
Adding to this debate, when referring to data-driven predictive approaches, we still keep the dichotomy between  direct/indirect approaches according to the following  mainstream terminology:
\begin{itemize}
    \item We call any control scheme that (directly) focuses on finding the control action as a function of data as \textbf{direct} (as exemplified by the straight arrow from the data to the controller in Fig. 1 of \cite{DorflerIEEECSM2023}). In the predictive control context, this is equivalent to focusing on estimating the control objective to be optimized. In probabilistic terms, this in turn requires \emph{finding the conditional distribution of the cost given observed data}, i.e., extracting from data no-more and no-less of what is needed for control design (a.k.a., a \textbf{minimal sufficient statistic for control}). Note that distributionally robust strategies such as \cite{CoulsonLD_DRO2022} fall with this class.    
    \item Conversely, an \textbf{indirect} approach undertakes intermediate steps that are \emph{not directly guided} by \emph{the control objective}. In the predictive control realm, these intermediate steps often coincide with the identification of a predictor, which is then fixed for control design. If the data are manipulated such that a sufficient statistic for control is not fully preserved and/or not properly exploited, this approach thus turns out to be sub-optimal. 
\end{itemize}
}


It is well-known that DeePC/DDPC, conventionally indicated as direct and model-free techniques, are equivalent to indirect and model-based (nominal) predictive control when measurements are noise-free  (see, e.g., \cite{breschi2022role}). Indeed, in a noise-free (deterministic) scenario, any finite segment of (informative) data is equivalent to having full knowledge of the model (or behavior) of the controlled system and, thus, of the cost. Therefore, in a deterministic (noise-free) context, classifying algorithms as direct/indirect and model-free/model-based is of purely nomenclature flavor.\\
{This is no longer true when data \emph{do not} follow exactly the laws of \emph{deterministic linear systems}. In this case, the distinctions between direct/indirect and model-based/model-free approaches become substantial, and the performance of ``deterministically inspired'' algorithms degrades quite rapidly \cite{dorfler2022bridging}. This can be interpreted as an overfitting phenomenon, in that the controller is overconfident in the estimate of the cost (or of the ``model/behavior'') and may attempt to compensate for noise in the training data, potentially destabilizing the closed loop system. Hence, accounting for uncertainties is of paramount importance, as already advocated (for model-based predictions) 
in the LQR scenario~\cite{Grimble:92}.}
{Recent research \cite{breschi2023impact,breschi2022uncertainty,CoulsonLD_DRO2022,dorfler2022bridging,Lazar2023} has focused on modifying DDPC algorithms exploiting regularization to make the optimal control input design robust against noise on the training data.  Simulation results show that regularization 
%
significantly improves control performance, when tuned correctly. However, tuning regularization penalties can be non-trivial,  requiring closed-loop experiments for validation \cite{dorfler2022bridging} or demanding the use of established rules of thumbs (see, e.g., \cite{Lazar2022}).} To cope with these limitations, the main contributions of this paper are as follows:
\begin{itemize}
\item  We introduce the \emph{Final  Control Error} (FCE) (see Definition \ref{def:FCE}), a ``control-oriented'' extension of the Final Prediction Error (FPE) \cite{akaikefpe}. The former provides an estimate of the actual (final) control cost based on the available (historical) data, becoming the objective to optimize when designing the predictive controller.
\item We show (see Section \ref{sec:SP}) that (i) the FCE can be ultimately decoupled as the sum of a \emph{certainty equivalence} (control) cost and a regularization term, with the latter quantifying the impact of uncertainty of the data-based predictions on the control objective, and (ii) that these two terms are functions of the predictor and its uncertainty. This result leads to a \emph{separation principle} for \emph{direct data-driven predictive control}, according to which 
the \emph{direct} predictive control design can be decoupled (or separated) in two steps: (i) compute an estimate of a predictor and its uncertainty, followed by the (ii) optimization of the FCE that is a function of the outcome of step (i). The practical byproduct of this is that of \emph{removing} the need for \emph{tuning} the regularization penalties, ultimately allowing to avoid lengthy or closed-loop calibration procedures. 
\item Our framework for direct data-driven predictive control that has \textit{the flexibility to incorporate  priors} on the  system to be controlled, a feature not readily available in existing  approaches. The non-informative prior scenario is derived as a special case (see Section \ref{Sec:NonInf}).
\item We demonstrate that existing predictive control schemes \cite{coulson2019data}, including recent variants like $\gamma$-DPPC \cite{breschi2022role}, can be viewed as \textit{sub-optimal instances of our formulation} (see Section \ref{sec:connections}).
\end{itemize}

Because of these results, our formulation  (see \textbf{Problem \ref{prob:opt:control_CLOSS}}) provides a \textbf{novel unified framework} for direct data-driven predictive control within a \emph{stochastic context}, where data are generated by an unknown linear time-invariant stochastic system. The resulting design procedure (see Section \ref{Sec:NonInf}) follows top-down without any additional assumption beyond linearity and time-invariance. Moreover, by establishing a \emph{separation principle} for predictive data-driven control, our formulation allows us to shed some light on the model-based vs model-free debate often addressed in the literature, clarifying the role that models and their uncertainty play in direct data-driven control. It should not come as a surprise that uncertainty plays a fundamental role in the data-driven control problem, as it does in any statistical decision problem.

The remainder of this paper unfolds as follows. Section~\ref{Sec:Problem Formulation} mathematically formulates the data-driven receding-horizon problem in the stochastic setting.  Section~\ref{Sec:HighLevelAss} discusses the minimal high-level assumptions necessary for formal analysis. In Section~\ref{sec:SP} a closed form expression for the Final Control Error  is provided and the \emph{separation principle} is stated.   Section~\ref{Sec:NonInf} \emph{specializes the results in the non-informative prior case}.  Existing methods are shown to be special cases within this framework in Section~\ref{sec:connections}. Numerical case studies are illustrated in Section~\ref{sec:examples}. The paper concludes with some final remarks (see Section~\ref{sec:conclusions}).

\textbf{Notation.} We denote with ${\mathbb N}$ the set of natural numbers (excluded zero) and ${\mathbb R}_+$ the set of non-negative real numbers. Given a matrix $A$, we denote its transpose as $A^{\top}$ and its Moore-Penroose pseudo-inverse as $A^{\dagger}$; $\|A\|$ and $\|A\|_F$ will denote respectively the $2$-norm and the Frobenius norm. In case $A$ is square, ${\rm diag}[A]$ indicates the diagonal matrix with diagonal elements equal to those in the diagonal of $A$ and, when $A$ is invertible, its inverse is denoted as $A^{-1}$. Meanwhile, we denote the identity matrix of size $d \times d$ as $I_{d}$. Given two vectors $a$ and $b$, the symbol $\aseq$ is used to denote their equality up to\footnote{
In asymptotic statistical analysis, it is often required that equality holds up to $o_{P}(1/\sqrt{N})$. Instead, here, we need equality up to $o_{P}(1/{N})$ since the cost is a quadratic function of the predictor and we need to make sure that certain 
approximations we introduce have an effect that can be neglected w.r.t. the dominating terms, which are of order $O(1/N)$.} $o_{P}(1/N)$, i.e.,   
\begin{equation}
a \aseq b \iff a=b+o_{P}(1/N),
\end{equation}
where the symbol $o_{P}(1/N)$ is defined as in \cite{breschi2022role,vanderVaart}. Given a sequence of random variables $x_N$ indexed by $N$  we denote with ${\mathrm{AsVar}}$ the variance of the limiting distribution as $N\rightarrow\infty$. Also, the symbol $\propto$ is used to indicate direct proportionality. For any signal $v(t) \in \R^v$, we define the associated Hankel matrix $V_{[t_0,t_1],N} \in \R^{v(t_1-t_0+1) \times N}$ as:
\begin{equation}\label{eq:Hankel}
V_{[t_0,t_1],N}\!:=\!\!\frac{1}{\sqrt{N}}\!\!\begin{bmatrix}
v(t_0) & v(t_0\!+\!1) \!&\! \cdots \!&\! v(t_0\!+\!N\!-\!1)\\
v(t_0\!+\!1) & v(t_0\!+\!2) \!&\! \cdots \!&\! v(t_0\!+\!N)\\
\vdots & \vdots \!&\! \ddots \!&\! \vdots\\
v(t_1) &v(t_1\!+\!1) \!&\! \dots \!&\! v(t_1\!+\!N\!-\!1)  
\end{bmatrix}\!\!,
\end{equation}
Finally, the shorthand $V_{t_0}:= V_{[t_0,t_0],N}$ is used to denote the Hankel containing a single row, namely
\begin{equation}\label{eq:Hankel:onerow}
V_{t_0}:= \frac{1}{\sqrt{N}}\begin{bmatrix}
v(t_0) & v(t_0\!+\!1) & \cdots & v(t_0\!+\!N\!-\!1)
\end{bmatrix}. 
\end{equation}

\section{From MPC to direct data-driven control}\label{Sec:Problem Formulation}

Consider a discrete-time, linear, time-invariant (LTI) system, with input $u(t) \in \mathbb{R}^{m}$ and output $y(t) \in \mathbb{R}^{p}$ and define the joint process
\begin{equation}\label{eq:z_def}
z(t):=\begin{bmatrix}
y(t)\\
u(t)
\end{bmatrix} \in \mathbb{R}^{m+p}.
\end{equation}
Assume that our goal is to control this system in a receding horizon fashion. 
Let us define the future input and output sequences
\begin{equation}\label{eq:future_input_optvar1}
\begin{array}{rcl}
u_f&:=&\begin{bmatrix}
u(t)^{\top} &
u(t\!+\!1)^{\top} &
\cdots &
u(t\!+\!T\!\!-\!1)^{\top}
\end{bmatrix}^{\top} \!\!\!
\\
y_{f}&:=&\begin{bmatrix}
y(t)^{\top} & y(t\!+\!1)^{\top} & \cdots & y(t\!+\!T\!-\!1)^{\top}
\end{bmatrix}^{\top} 
\end{array}
\end{equation}
and the following collection of $\rho$ joint inputs and outputs:
\begin{equation}\label{eq:zinit}		
z_{ini}:=\begin{bmatrix}
z(t\!-\!\rho)^\top & 
\cdots &
z(t\!-\!2)^\top & 
z(t\!-\!1)^\top
\end{bmatrix}^\top\!\!,
\end{equation}
encoding the initial condition of the controlled system at time $t$.
In model predictive control a central role is played by the ``true'' model ${\mathcal{M}}$ and the corresponding  (model-based) output predictor, i.e.: 
\begin{equation}\label{eq:multistep_predictor}
\bar y_f(u_f): = {\mathbb E}[y_f|z_{ini},u_f,{\mathcal{M}}].
\end{equation}

At each time step $t \in \mathbb{N}$, we aim at solving the optimal predictive control problem
\setcounter{problem}{-1}
\begin{problem}[Model Predictive Control]\label{prob:MPC}
\begin{equation}\label{eq:opt:control}
u_f^\star: = \mathop{\rm arg\min}_{u_f \in {\mathcal U}_f, \bar y_f \in {\mathcal Y}_f}~L_t(u_f)
\end{equation} 
\end{problem}
seeking for a feasible future input sequence
that minimizes a user-defined time-dependent cost $L_{t}\!:\! \mathbb{R}^{\!mT} \rightarrow \mathbb{R}$ over a given horizon of length $T \geq 1$. Computing the cost $L_{t}(u_f)$, that needs to be updated at each control step $t$, requires knowledge of the \textquotedblleft true\textquotedblright \ data-generating mechanism\footnote{Because of this dependence, we refer to \eqref{eq:opt:control} as the optimal, model-based problem.} $\mathcal{M}$. Only the first optimal control\footnote{
In principle one could apply the first $s\leq T$ computed optimal controls, leading to a $s-$step predictive control scheme.}  action $u^{\star}(t)$ is then applied to the actual system, while the remaining optimal inputs are discarded, and \eqref{eq:opt:control} is solved again at the next step $t+1$.

In this paper, we work with a quadratic cost $L_t(u_f)$ like in standard MPC \cite{borrelli2017predictive} and existing data-driven predictive control methods \cite{berberich2020data,breschi2022role,Coulson2019}. In particular, $L_t(u_f)$ will be a function of the input and output references $u_r \in \mathcal{U}_{f}^{T}$ and $y_r \in \mathcal{Y}_{f}^{T}$ as follows:
\begin{align}\label{eq:MP_cost}
\nonumber L_t(u_f)&\!=\! \E\!\left[ \| y_r \!-\! y_f\|_{Q_o}^2 \!+\! \| u_r \!-\! u_f\|_{R}^2| {\mathcal M} \right] \\
\nonumber &\!=\! 
\| y_r \!-\! \bar y_f(u_f)\|_{Q_o}^2 \!+\! \| u_r \!-\! u_f\|_{R}^2 \\
\nonumber &~~~\!+\!\E\!\left[ \| y_f \!-\! \bar y_f(u_f)\|_{Q_o}^2\right] \\
&\!\propto\!  \| y_r \!-\! \bar y_f(u_f)\|_{Q_o}^2 \!+\! \| u_r \!-\! u_f\|_{R}^2,
\end{align}
where $Q_o \in \mathbb{R}^{pT \times pT}$ and $R \in \mathbb{R}^{mT \times mT}$ are positive definite matrices to be freely chosen by the control designer to achieve specific goals, while $\E\!\left[ \| y_f \!-\! \bar y_f(u_f)\|_{Q_o}^2\right]$ can be neglected since it only involves expected value of future noise but it does not depend on $u_f$ (see \cite{breschi2022role}). Note that, the conditional expectation on $\mathcal{M}$ implies having \textquotedblleft full knowledge\textquotedblright \ of the system to be controlled, i.e., knowing its model (and, thus, the functional form of the predictor) and having 
perfect knowledge of noise statistics, as typical in (deterministic and stochastic) MPC (see \cite{borrelli2017predictive}). 

On the wave of data-driven predictive control approaches and, thus, differently from MPC, in this work we are interested in the case where the predictor $ \bar y_f(u_f)$ for the system is \emph{unknown}, while we have access to a finite set of input/output (training) data ${\mathcal D}\!:=\!\{(y(s),u(s)), s\!=\!1,\ldots,N_{data}\}$ collected in open loop (namely, before closed-loop operations are initiated). In this context, the sufficient statistic for solving the targeted optimal control problem \eqref{eq:opt:control} is the 
conditional probability:
\begin{equation}\label{eq:prob:cost}
P(L|u_f,{\mathcal D}) := {\mathbb P}[L_t(u_f) \leq L | {\mathcal D}].
\end{equation}

Following the terminology used in \cite{akaikefpe}, where the term \emph{Final Prediction Error} (FPE) is used to denote the final (i.e., accounting for model uncertainty) mean squared error obtained when predicting future data based on a predictor estimated from historical data, we shall give the following definition: 

\begin{definition}\label{def:FCE}
The conditional mean ${\mathbb E}[L_t(u_f)|{\mathcal D}]$, i.e., the estimate of the (finite-horizon, open loop) control cost based on the training data ${\mathcal D}$, will be called \emph{Final Control Error} (FCE).
\end{definition}

The FCE is a function of the control actions (that we need to optimize) and  of the available (training) data ${\mathcal D}.$ Accordingly, we recast the design problem in \eqref{eq:opt:control} as the optimization of the Final Control Error, i.e.,\vspace{.1cm}
\begin{problem}[FCE-DDPC]\label{prob:opt:control_CLOSS}
\begin{equation}\label{eq:opt:control_CLOSS}
\hat u_f^\star: = \mathop{\rm arg\min}_{u_f \in {\mathcal U}_f, {\mathbb E[\bar y_f|{\mathcal D}] \in {\mathcal Y}_f}}~\underbrace{{\mathbb E}[L_t(u_f)|{\mathcal D}]}_{:=FCE(u_f)},
\end{equation}
\end{problem}
whose solution $\hat u_f^\star$ of \eqref{eq:opt:control_CLOSS} is a \emph{measurable function of the data} $\mathcal{D}$, whereas $u_f^\star$ in \eqref{eq:opt:control} is a deterministic quantity.
\begin{remark}[Alternative formulations]
Based on the conditional probability in \eqref{eq:prob:cost} different notions of optimality for control can be considered, leading to alternative reformulations of the design problem in \eqref{eq:opt:control}. A first possibility would be to minimize the so-called Value at Risk (VaR) \cite{ArtznerCvar1999} for a given confidence level $\alpha$. This would result in the following problem:
\begin{align*}\label{eq:opt:control_VAR}
\hat u_f^\star: = &\mathop{\rm arg\min}_{u_f \in {\mathcal U}_f, {\mathbb E}[\bar y_f|{\mathcal D}] \in {\mathcal Y}_f,\ell}~\ell\\
& \qquad {\rm{s.t. }}~~{\mathbb P}[L_t(u_f) > \ell | {\mathcal D}] = \alpha. 
\end{align*}
A more cautious alternative would be to optimize the so-called Conditional Value at Risk (CVaR) \cite{ArtznerCvar1999}, i.e., to solve
\begin{equation}\label{eq:opt:control_CVAR}
\begin{aligned}
\hat u_f^\star: = &\mathop{\rm arg\min}_{u_f \in {\mathcal U}_f, {\mathbb E}[\bar y_f|{\mathcal D}] \in {\mathcal Y}_f,\textcolor{black}{\ell}}~{\mathbb E}[L_t(u_f)| L_t(u_f) > \ell, {\mathcal D}]\\
& \qquad {\rm{s.t. }}~~{\mathbb P}[L_t(u_f) > \ell | {\mathcal D}] = \alpha. 
\end{aligned}
\end{equation}
In this case, one is thus interested not only in bounding the probability of losses exceeding a certain threshold (as for the VaR), but also in limiting the expected loss in case such threshold is exceeded. While we defer an analysis of the implications that such changes in the notion of optimality have on the final closed-loop performance to future work, we wish to stress that the considered \textbf{Problem~\ref{prob:opt:control_CLOSS}} in \eqref{eq:opt:control_CLOSS} corresponds to \eqref{eq:opt:control_CVAR} for $\alpha=1$.
\end{remark}

\section{High level assumptions on the predictor}\label{Sec:HighLevelAss}
Given the input-output measurements available up to time $t$, here denoted as  $z_{t}^{\!-}\!=\!\{z(t-k)\}_{k \in \mathbb{N}}$, let us consider the one-step predictor
\begin{equation}\label{eq:general_predictor}
\bar{y}(t|t-1):=\mathbb{E}[y(t)|z_{t}^{\!-},{\mathcal M}],
\end{equation} 
on which we make the following assumptions.

\begin{assumption}\label{ass:lin_pred}
The predictor \eqref{eq:general_predictor} is linear and has an $\ell_{1}$ (i.e., BIBO) stable impulse response, namely: 
\begin{align}\label{eq:predictor}
\nonumber \bar y(t|t-1) &= \sum_{k=1}^{+\infty} \phi_{k} z(t-k), \\
& = \sum_{k=1}^{+\infty} \phi^y_{k} y(t-k) +  \phi^u_{k} u(t-k),
\end{align}
with $\{\phi_{k}\}_{k \in \mathbb{N}} \in \ell_{1}$ and $\phi_{k}:=\begin{smallmatrix}\begin{bmatrix}
		\phi_{k}^{y} & \phi_{k}^{u}
\end{bmatrix}\end{smallmatrix}$.
\end{assumption}
\begin{assumption}\label{ass:Martingale}
The one-step prediction error $e(t):=y(t)-\bar{y}(t|t-1)$ is a martingale difference with constant conditional variance $\sigma^{2}I_{p}$
, namely:
\begin{subequations}\label{eq:features_error}
\begin{align}
& \mathbb{E}[e(t)|z_{t}^{\!-},u_t^+]=0\\ 
& {\mathrm{Var}}[e(t)|z_{t}^{\!-},u_t^+]={\mathrm{Var}}[e(t)]=\sigma^{2}I_{p},
\end{align}
\end{subequations}
conditionally on the joint input-output past data $z_{t}^{\!-}$, and future inputs $u_{t}^{\!+}:=\{u(t+k-1)\}_{k \in {\mathbb N}}$. 
\end{assumption}
While Assumption~\ref{ass:lin_pred} implies that $y(t)$ admits an \emph{infinite order ARX} representation with BIBO-stable impulse response, Assumption~\ref{ass:Martingale} requires that data are gathered in open loop (the reader is referred to \cite{ChiusoMoffatDorfler2023}, where the closed-loop case is treated). 

\begin{remark}[On the role of \eqref{eq:predictor}]
The representation in \eqref{eq:predictor} should not be regarded as a tentative parameterization of the predictor via the (infinite dimensional) set of ``parameters''  $\Phi:= \{\phi_k\}_{k\in {\mathbb N}}$, but rather as a general representation encompassing all  linear and bounded maps between past inputs/outputs and $\bar{y}(t|t-1)$. Indeed, our assumptions are rather mild and not limiting, as the considered class of predictors spans all linear (and even infinite-dimensional), 
bounded operators mapping past data into the one-step-ahead output prediction. 
\end{remark}  


Since $\{\phi_{k}\}_{k \in \mathbb{N}} \in \ell_1$, it is  possible to truncate the 
the predictor in \eqref{eq:predictor} so that the difference
\begin{equation}\label{eq:predictor_err}
\bar y(t|t-1) - \left[\sum_{k=1}^{\rho} \phi^y_{k} y(t-k) + \phi^u_{k} u(t-k)\right]
\end{equation}
can be made arbitrarily small provided $\rho$ is large enough \cite{Bauer-05,Bauer-L-01,CHIUSOvecRegModeling2007}.

\begin{remark}[On the choice of $\rho$]
The approximation error in \eqref{eq:predictor_err} is determined by ${\rho}$ and the true
model. Nonetheless, there exist data-driven selection rules $\hat\rho$, e.g. based on AIC \cite{Bauer-05}, such that the difference in \eqref{eq:predictor_err}  is  $o(1/N_{data})$, i.e.,
\begin{equation}\label{eq:predictor2}
\bar y(t|t-1) \doteq \sum_{k=1}^{\hat\rho} \phi^y_{k} y(t-k) + \phi^u_{k} u(t-k).
\end{equation}
 
\end{remark}

Via the truncated representation in \eqref{eq:predictor2}, and defining
\begin{align}\label{eq:useful_matrices}
\nonumber & \Phi_{\!u} \!\!=\!\!\! \begin{bmatrix} 
\!0 &\! 0  & 0 \!&\! \dots \!& 0  \\
\!\phi_{1}^u &\! 0  &  0 \!&\! \dots \!& 0 \\
\!\phi_2^u  &\! \phi_1^u  & 0  \!&\!  \dots \!& 0 \\
\!\vdots &\! \vdots  \!&\! \ddots \!&\!  \ddots \!& \vdots \\
\!\phi_{T\!-\!1}^u  &\! \phi_{T\!-\!2}^u   &\! \dots  \!&\! \phi_1^u \!& 0
\end{bmatrix}\!\!,~~
\Phi_{\!y} \!\!=\!\!\! \begin{bmatrix} 
\!0 &\! 0  & 0 \!&\! \dots \!& 0  \\
\!\phi_1^y &\! 0  &  0 \!&\! \dots \!& 0 \\
\!\phi_2^y  &\! \phi_1^y  & 0  \!&\!  \dots \!& 0 \\
\!\vdots &\! \vdots  \!&\! \ddots \!&\!  \ddots \!& \vdots \\
\!\phi_{T\!-\!1}^y  &\! \phi_{T\!-\!2}^y   &\! \dots  \!&\! \phi_1^y \!& 0
\end{bmatrix}\!\!,\\
& \qquad \Phi_P = \begin{bmatrix} 
\phi_\rho &\phi_{\rho-1}  &  \phi_{\rho-2} & \dots & \dots & \dots & \phi_{1}  \\
0 & \phi_\rho  &  \phi_{\rho-1} & \dots& \dots&\dots &  \phi_{2} \\
0 & 0  &  \phi_\rho  &  \dots & \dots& \dots&   \phi_{3}  \\
\vdots & \vdots  & \ddots &  \ddots & \vdots& \vdots& \vdots   \\
0  & 0   & \dots  & 0 & \phi_\rho &\dots& \phi_{T}
\end{bmatrix} ,
\end{align}
\begin{equation}\label{eq:model_basedweight}
	W:=I_{pT}-\Phi_{y},
\end{equation}
\begin{equation}\label{eq:future_input_optvar}
\begin{array}{rcl}
e_f&:=&\begin{bmatrix}\nonumber
e(t)^{\top} &
e(t\!+\!1)^{\top} &
\cdots &
e(t\!+\!T\!\!-\!1)^{\top}
\end{bmatrix}^{\top}, \!\!\!
\end{array}
\end{equation}
we can now formally characterize the multi-step predictor as follows. 
\begin{proposition}\label{prop:opt:pred}
Recalling the definitions of $u_{f}$, $y_{f}$, and $z_{ini}$ in \eqref{eq:future_input_optvar1} and \eqref{eq:zinit}, the optimal $T$-step-ahead output predictor $\bar y_f(u_f)$ satisfies 
\begin{subequations}
\begin{equation}\label{eq:predictor3}
W \bar y_f(u_f)~\aseq~\Phi_P z_{ini} + \Phi_u u_f,
\end{equation}
and
\begin{equation}\label{eq:predictor_Error}
y_f- \bar y_f(u_f)~\aseq~W^{-1} e_f.
\end{equation}
\end{subequations}
\end{proposition}
\begin{proof}
See the Appendix.
\end{proof}
\section{A Separation Principle for DDPC} \label{sec:SP}

We now establish a \emph{separation principle} for  DDPC, by showing that one needs estimates of the model together with its uncertainty to compute the cost to be optimized (i.e., the Final Control Error).

To streamline notation, let us denote the difference between the predicted tracking error as 
\begin{equation}\label{eq:tackingerror}
\delta(u_f)\!=\!y_{r}-\bar y_f(u_{f}),
\end{equation}
so that the cost \eqref{eq:MP_cost} can be compactly rewritten as  
\begin{equation*}
	L_{t}(u_{f}) = \|\delta(u_f)\|_{Q_o}^{2}\!\!+\|u_{r}-u_{f}\|_{R}^{2}.
\end{equation*}
Based on \eqref{eq:predictor_Error},
it is straightforward to prove that the $T$-step-ahead prediction error covariance is given by
$\sigma^2 W^{-1}W^{-\top}$. We can account for this by redefining the tracking loss as follows:
\begin{equation}\label{eq:costdet}
	L_{t}(u_{f}):=\|\delta_W(u_f)\|_{Q}^{2}+\|u_{r}-u_{f}\|_{R}^{2},
\end{equation}
where $Q:=W^{-\top}Q_oW^{-1}$  and
\begin{equation}\label{eq:deltaW}
\delta_{W}(u_f):=W\delta(u_f) 
\doteq y_r - \Phi_y y_r - \Phi_P z_{ini} - \Phi_u u_f,
\end{equation} 
with the  last equality stemming form \eqref{eq:predictor3} and \eqref{eq:tackingerror}.
To simplify the notation, the explicit dependence of $\delta_W$ on $u_f$ will be omitted in what follows.

To re-write \eqref{eq:deltaW} in a way that is more amenable to computations, we now introduce the matrix 
\begin{equation}\label{eq:Theta}
\Theta: = \begin{bmatrix}
 \phi_1 & \phi_2 & \cdots &\phi_\rho \end{bmatrix}
\end{equation} and its vectorization  $\theta: = {\rm vec}[\Theta] \in \mathbb{R}^{p(m+p)\rho}$, which fully  characterizes the predictor in \eqref{eq:predictor3}. Note now  that $\delta_W(u_f)$ in \eqref{eq:deltaW} can be written as the sum of three terms:  $y_r$; a linear term in $\theta$; and one  bilinear in $\theta$  and $u_f$. Therefore we can find a constant selection matrix $M_0  \in \mathbb{R}^{pT \times p(m+p)\rho}$ and a selection matrix $M(u_f) \in \mathbb{R}^{pT \times p(m+p)\rho}$ (linear in $u_f$) such that:
\begin{equation}
\label{eq:lin_param}
\delta_W(u_f):=
y_r - [M_0 + M(u_{f})]\theta.
\end{equation}
For future use, we then introduce the following notations for the conditional mean  and conditional variance of $\theta$:
\begin{equation}\label{eq:param_conditional}
\bar{\theta} := \mathbb{E}[\theta|\mathcal{D}] \quad \quad \Sigma_{\theta}:=Var[\theta|\mathcal{D}].
\end{equation}
Exploiting this notation, we provide an explicit expression for the conditional loss in the following theorem, that can be interpreted as a \emph{separation principle} for data-driven predictive control. Indeed, it highlights that the cost to be optimized is a measurable function of the (training) data only through the conditional mean and variance \eqref{eq:param_conditional} of the predictor coefficients. In turn, this shows that the estimate of a (non-parametric) \emph{model} and its uncertainty are (Bayesian) \emph{sufficient statistics} for control.  
\begin{theorem}\label{prop:condcost}
{\bf (Separation Principle)} Let $L_t(u_f)$ be defined as in \eqref{eq:costdet}, the Final Control Error in \eqref{eq:opt:control_CLOSS} is given by
\begin{subequations}\label{eq:condLossW}
\begin{equation}
FCE(u_f) = \mathbb{E}[L_{t}(u_{f})|\mathcal{D}]\doteq  J(u_{f})+r(u_{f}),
\end{equation}	
where
\begin{align}
&J(u_{f}):= \|\overline{\delta}_W(u_f)\|_Q^{2}+\|u_{r}-u_{f}\|_{R}^{2},\label{eq:certainty_1}\\
& r(u_{f}):={\mathrm{Tr}} \left[Q{\mathrm{Var}}[\delta_W(u_f)|\mathcal{D}]\right],\label{eq:reg_1}
\end{align}
and 
\begin{equation}\label{eq:exp_weighted_error}
\begin{aligned}
&\overline{\delta}_W(u_f):=\mathbb{E}[\delta_W(u_f)|\mathcal{D}]= y_r- [M_0 + M(u_f)]\bar \theta, \\
&{\mathrm{Var}}\left[\delta_W(u_f)|\mathcal{D}]\right]= \left(M_0\!+\!M(u_f)\right) \Sigma_{\theta} \left(M_0\!+\!M(u_f)\right) ^\top\!\!.
\end{aligned}
\end{equation}
\end{subequations} 
\end{theorem}
\begin{proof}
The proof follows immediately from the decomposition of the conditional expected value in \eqref{eq:costdet} in terms of ``squared mean plus variance''. Indeed, given the conditional mean defined as in \eqref{eq:exp_weighted_error}, \eqref{eq:costdet} can be rewritten as 
$$
\begin{array}{rcl}
L_t(u_f) & = & \|\delta_W(u_f)\|^2_Q +  \|u_r - u_f\|_R^2 \\
& = & \|\overline{\delta}_W(u_f)\|^2_Q + \|u_r - u_f\|_R^2 \\
& & + \|\delta_W(u_f)-\overline{\delta}_W(u_f)\|_Q^2\\
& & + 2 {\mathrm{Tr}}\left[Q (\delta_W(u_f)-\overline{\delta}_W(u_f))\overline{\delta}_W(u_f)^\top\right].
\end{array}$$
Using the facts that $\overline{\delta}_W(u_f)$ is measurable w.r.t. ${\mathcal D}$, $\E[\delta_W(u_f)-\overline{\delta}_W(u_f)|{\mathcal D}]=0$ and that
$$
\E\left[\|\delta_W(u_f)-\overline{\delta}_W(u_f)\|_Q^2|{\mathcal D} \right] = {\mathrm{Tr}}\left[Q {\mathrm{Var}}\left[\delta_W(u_f) |{\mathcal D}\right]\right],
$$
Finally, it is sufficient to recall that $\delta_W(u_f) $ is a linear function of $\theta$ (see \eqref{eq:lin_param}), from which the conclusion follows. 
\end{proof}


Note that $J(u_f)$ and $r(u_f)$  given in Theorem \ref{prop:condcost}
can be respectively regarded as the \textquotedblleft certainty equivalence quadratic\textquotedblright \ loss, and a regularization that controls the effects of uncertainties on the future output predictions. 

\begin{remark}\label{rem:Q}
The cost in \eqref{eq:condLossW} still depends on the choice of the weightings $Q$ and $R$. These degrees of freedom can be exploited to capture specific problems encountered in the literature in designing data-driven predictive controllers. For instance, defining $\bar W$ by replacing the unknown $\phi$'s in \eqref{eq:useful_matrices}, \eqref{eq:model_basedweight} with their estimates contained in the vector $\bar \theta$ and assigning to $Q$ the value $\overline Q:=\overline W^{-\top}Q_o \overline W^{-1} $ 
we obtain
\begin{align}
\nonumber & \bar{J}(u_f):=\|\overline{\delta}_W(u_f)\|_{\overline Q}^{2}+\|u_r-u_f\|_{R}^{2}\label{eq:DDDPCP_1J_FINAL}\\
&\qquad =\|y_r\!-\!\bar W^{-1}\mathbb{E}[W\bar y_f(u_f)|{\mathcal D}]\|_{Q_o }^{2}\!\!+\!\|u_r\!-\!u_f\|_{R}^{2},\\
&\bar{r}(u_f):={\mathrm{Tr}} \left[\overline{W}^{-\top}Q_o \overline{W}^{-1}\, {\mathrm{Var}}[\delta_W(u_f)|\mathcal{D}]\right],\label{eq:DDDPCP_1r_FINAL}
\end{align}
so that $\bar{J}(u_f)$ is the control cost in \cite{breschi2022design,dorfler2022bridging} and $\bar{r}(u_f)$ is the corresponding  regularizer. 
\end{remark}
\subsection{On the model-based vs model-free dichotomy}
The recent literature has focused on the role that models play in data-driven control approaches, see e.g., \cite{DorflerIEEECSM2023} discussing pros and cons of model-based vs model-free approaches. 
 Restricting the attention to the LTI setup, 
Theorem \ref{prop:condcost} sheds some new light on the model-free vs model-based dichotomy. Specifically, our \emph{separation principle} implies that, under the sole LTI assumption on the underlying plant, a non-parametric model (a near-optimal restricted complexity model  \cite{Hjalmasson-05}) is always behind the curtain. Indeed, to compute the FCE (i.e., a data-driven estimate of the objective function) one needs to compute an estimate of the predictive model together with its uncertainty. \\
This is supported by the results in Section \ref{sec:connections}, showing that popular approaches such as DeePC and $\gamma$-DDPC 
can be seen as special cases of Theorem~\ref{prop:condcost}.\\
As such, we feel the model-free vs. model-based dichotomy should be down-emphasized. Rather, we would argue that approaches that are typically referred to as \emph{model-based} just entail further \emph{restrictions} to the model class beyond linearity, e.g., identifying an LTI state-space model of fixed order. The Bayesian approach proposed in this paper includes all these as special cases by fixing the  prior to be the indicator function of a specific model class.
\section{The non-informative prior case }\label{Sec:NonInf}
In this Section, we specialize the solution provided in Theorem \ref{prop:condcost} when no prior assumptions are made on the model except for linearity, which is common to DDPC approaches (see e.g., \cite{berberich2020data,breschi2022design,dorfler2022bridging}). To do so, we need to first postulate that the prior on the predictor coefficients is Gaussian, thus yielding closed-form expressions for the conditional mean and variance in \eqref{eq:param_conditional}, and then consider the non-informative prior limit as explained in what follows. 

In this Section, we then work under the following assumption.
\begin{assumption}\label{ass:GaussianPrior}
   The sequence of the predictor coefficients $\{\phi_{k}\}_{k \in \mathbb{N}}$ is a zero-mean, Gaussian process
\begin{equation}\label{eq:GP:prior}
\begin{aligned}
&\{\phi_{k}^{u}\}_{k \in \mathbb{N}} \sim \mathcal{GP}(0,K_{\lambda}),~~~ \{\phi_{k}^{y}\}_{k \in \mathbb{N}} \sim \mathcal{GP}(0,K_{\lambda})\\
& \qquad \qquad \qquad \{\phi_{k}^{u}\}_{k \in \mathbb{N}} \perp \{\phi_{k}^{y}\}_{k \in \mathbb{N}},
\end{aligned}
\end{equation}
with covariance $K_{\lambda} := \lambda K$, and where the kernel $K$ only has to result in $\ell_1$-stable realizations of this prior.
\end{assumption}

The \emph{non-informative} prior limit will be obtained by letting the scaling factor on the prior variance to go to infinity, i.e., $\lambda \rightarrow \infty$.
  
Let us also define the block Hankel matrices:
\begin{equation}\label{eq:past_dataMatrix}
Z_{P}:=Z_{[1,{\rho}],N},
\end{equation}
and
\begin{equation}\label{eq:outputMatrix}
Y_{{\rho}+1}:=Y_{[{\rho}+1,{\rho}+1],N},
\end{equation}
with $N=N_{data}-{\rho}$.
We are now ready to state the main results of this Section.
\begin{lemma}\label{lem:main}
Assume that the one step ahead prediction error is Gaussian, the prior for the data generating mechanism is given in  \eqref{eq:GP:prior} with $\lambda \rightarrow \infty$. Under the approximation that the data $\{z(s)\}_{s=1}^{\rho}$ are only exploited to fix the initial condition (see \cite[Proposition 3 and equation (34)]{PILLONETTO2011291}) the following holds:
\begin{align}
\nonumber &\bar{\theta}= {\rm vec}\left[Y_{\rho+1}Z_{P}^{\!\top}(Z_{P}Z_{P}^{\!\top})^{-1}\right]\\
&~~ = \left[\left(Z_P Z_P^\top\right)^{-1} \otimes I_p\right]\left(Z_P \otimes I_p\right){\rm vec}\left(Y_{\rho+1}\right), \label{eq:hat_theta} \\
&\Sigma_{\theta}=\sigma^{2}\left[(Z_{P}Z_{P}^{\!\top})^{-1} \otimes I_p\right]. \label{eq:sigma_theta}
\end{align}
\end{lemma}
\begin{proof}
Let us first note that exploiting the expression for the one-step-ahead predictor in \eqref{eq:predictor2} the data Hankel  matrices satisfy
\begin{equation}\label{eq:ARX:param}
Y_{\rho+1}\aseq \Theta Z_P + E_{\rho+1}.
\end{equation} 
Therefore, defining $\bar y:= {\rm vec}\left[Y_{\rho+1}\right]$ and $\bar e:= {\rm vec}\left[E_{\rho+1}\right]$, \eqref{eq:ARX:param}  can be vectorized, yielding to:
$$
\bar y = (Z_P^\top \otimes I_p) \theta + \bar e$$
Under the Gaussian prior \eqref{eq:GP:prior} for the predictor coefficients, there exists a  symmetric positive definite matrix $P$ such that 
$\theta \sim {\mathcal N}(0,\lambda P)$, so that 
the conditional mean and conditional variance are given by 
\cite{Book_RegID2022}:
\begin{subequations}
\begin{align}
& \E[\theta | {\mathcal D}] \!\aseq\!\! \left[\left(Z_P Z_P^\top \otimes I_p\right) \!+\! \frac{\sigma^2}{\lambda} P^{-1}\right]^{-1\!}\!\!\!\left(Z_P \otimes I_p\right)\bar y, 
\label{eq:cond_mean} \\
& {\mathrm{Var}}[\theta | {\mathcal D}] \aseq \sigma^2 \left[\left(Z_P Z_P^\top\otimes I_p\right) + \frac{P^{-1}}{\lambda} \right]^{-1},  \label{eq:cond_var}
\end{align}
\end{subequations}
in turn obtained based on the following property of Kronecker products:
$$
\left(Z_P\otimes I_p\right)\left(Z_P  \otimes I_p\right)^\top=\left(Z_P Z_P^\top \otimes I_p\right).
$$
Taking the limit as $\lambda \rightarrow \infty$ in \eqref{eq:cond_mean}, \eqref{eq:cond_var} and exploiting the 
identity
$$
\left(Z_P Z_P^\top \otimes I_p\right)^{-1} =\left(Z_P Z_P^\top\right)^{-1} \otimes I_p,
$$
\eqref{eq:hat_theta} and \eqref{eq:sigma_theta} follow.
\end{proof}

The reader may observe that \eqref{eq:hat_theta} and \eqref{eq:sigma_theta} are the least squares estimate of the ARX coefficients $\theta$ and their variance.
\begin{remark}[Truncation in a Bayesian setup]\label{rem:GP}
	For finite $\lambda$ in \eqref{eq:GP:prior}, the truncation to $\hat{\rho}$ introduced in \eqref{eq:predictor2} can actually be avoided, since the posterior means and variances \eqref{eq:cond_mean}, \eqref{eq:cond_var}  are well-defined (i.e., bounded) even for $\hat{\rho} \rightarrow \infty$. We leave the analysis of this scenario to future work.
\end{remark}

%
%
Lemma~\ref{lem:main} ultimately allows us to obtain a computable expression for the Final Control Error, as  follows.
\begin{theorem}[FCE with non-informative prior]\label{thm:FCE:NIP}
Under the assumptions of Lemma~\ref{lem:main}, the Final Control Error optimized in \textbf{Problem~\ref{prob:opt:control_CLOSS}} admits the expression: 
\begin{subequations}\label{eq:FCE:NIP}
	\begin{equation}
		FCE(u_f) = J(u_f)+r(u_f),
	\end{equation}
where
\begin{align}
	& J(u_f)\!=\!\|y_r-(M_{0}+M(u_f))\bar{\theta}\|_{Q}^{2}+\|u_{r}-u_{f}\|_{R}^{2}, \label{eq:opt:cost}\\
	& r(u_f)\!=\!\mathrm{Tr}\! \left[Q(M_{0}+M(u_f))\Sigma_{\theta}(M_{0}+M(u_f))^{\!\!\top}\right], \label{eq:opt:regularizer}
\end{align}	
\end{subequations}
and $\bar{\theta}$ and $\Sigma_{\theta}$ are the data-driven quantities defined as in \eqref{eq:hat_theta} and \eqref{eq:sigma_theta}, respectively.
\end{theorem}

Let us recall that $M(u_f)$ in \eqref{eq:opt:cost} and \eqref{eq:opt:regularizer} is linear in $u_f$. Therefore, the problem in \eqref{eq:opt:control_CLOSS} is quadratic in $u_{f}$, reflecting the presence of 
a quadratic regularization term in most of the existing formulations of data-driven predictive control, see e.g., \cite{berberich2020data,breschi2023impact,dorfler2022bridging}. At the same time, differently from existing DDPC problems, the one we obtain has important properties.
\begin{enumerate}
\item The  regularizer \eqref{eq:opt:regularizer} \textbf{depends explicitly} on the training \textbf{data} and on the \textbf{reference} trajectory. This feature is not shared by most regularization schemes, where offline tuning of the regularization penalties partially compensates for this. 
\item  \textbf{No tuning} of regularization parameter is needed, as the value of this penalty comes directly from \eqref{eq:sigma_theta}.
\end{enumerate}
This last statement stems from the following remark.
\begin{remark}[Dependence of \eqref{eq:opt:regularizer} on $\sigma^{2}$]\label{remark:sigma}
	Based on \eqref{eq:sigma_theta} the regularization term in \eqref{eq:opt:regularizer} depends on the (unknown) conditional variance of the one-step prediction error \eqref{eq:features_error}. This quantity can be estimated from $\mathcal{D}$ as the sample variance of the ARX model residuals 
	$$
	\hat \sigma^2:= \frac{1}{p(N-d)} \left\| Y_{\rho + 1}\left(I - Z_P^\top(Z_P Z_P^\top\right)^{-1}Z_P)\right\|_F ^2,
	$$
	where $d = (m+p)\rho$ is the number of free parameters  for each output. 
\end{remark}
Two additional remarks are also in order.
\begin{remark}[Sufficient Statistics for  control]\label{remark:SufficiantStatistics}
The optimization problem \eqref{eq:opt:control_CLOSS} with $J(u_f)$ and $r(u_f)$ as in \eqref{eq:opt:cost}, \eqref{eq:opt:regularizer} depend on the data ${\mathcal D}$ (see \eqref{eq:hat_theta} and \eqref{eq:sigma_theta}) through the sufficient statistics
\begin{equation}\label{eq:suffstat}
Y_{\rho+1}Z_{P}^{\!\top}, \quad \quad Z_{P}Z_{P}^{\!\top},
\end{equation}
which are 
sample (second) moments of the measured data. It is worth observing that \eqref{eq:suffstat} coincide with the 
sufficient statistics for the coefficients of the predictor \eqref{eq:predictor}.  The function that maps the sufficient statistics to the FCE is not a bijection, and thus a lower dimensional (minimal) sufficient statistic may be 
found. 
However, this function that ``compresses'' the sufficient statistic depends upon the initial conditions $z_{ini}$ and reference signals $y_r$ and $u_r$. Since we are interested in optimizing \eqref{eq:condLossW} for any possible choice of $z_{ini}$, $y_r$ and $u_r$,  \eqref{eq:suffstat} is indeed the minimal information of the data one needs to store.
\end{remark}

\begin{remark}[Exploiting Informative Priors]\label{remark:prior} The proposed setup allows us to employ informative priors (e.g., stable and smoothness priors such as TC/DC \cite{Book_RegID2022}) 
by just exploiting the expressions \eqref{eq:cond_mean} and \eqref{eq:cond_var} to compute the cost in \textbf{Problem~\ref{prob:opt:control_CLOSS}}, which is not as straightforward in other DDPC approaches. This is the reason why we have decided to follow a Bayesian viewpoint to obtain the ``frequentist'' case as the non-informative prior limit.
We should like to stress that also classical Model Predictive Control (see \textbf{Problem~\ref{prob:MPC}}), in which one has full knowledge of the system's model, can be obtained as the limit of a point (delta) prior centered on the \textquotedblleft true\textquotedblright \ model ${\mathcal M}$.
\end{remark}


\section{Connection with DeePC and $\gamma$-DDPC}\label{sec:connections}
In this Section, we focus on highlighting how two recently proposed techniques for data-driven predictive control, namely the DeePC scheme with \textquotedblleft consistency regularizer\textquotedblright \ proposed in \cite[Section IV.B]{dorfler2022bridging} and $\gamma$-DDPC \cite{breschi2022role}, fit within the framework of \textbf{Problem~\ref{prob:opt:control_CLOSS}}. To this purpose, 
we introduce some preliminary definitions and notation.  

First of all let us introduce the Hankel matrices
\footnote{For $N_{data}$ training data, the Hankel data matrices $Z_P,Y_F$ and $U_F$ must have $N=N_{data}-\rho-T+1$ columns.}:
\begin{subequations}\label{eq:Hankel_data}
\begin{align}
&U_{F}:=U_{[\rho,\rho+T-1],N},~~Y_{F}:=Y_{[\rho,\rho+T-1],N},\\
& H_{ZU}:=\left[\begin{matrix} Z_P \\ U_F\end{matrix}\right], 
\end{align}
\end{subequations}
respectively collecting ``future''  data and stacking  ``past'' joint trajectories and  ``future'' outputs, and let us denote the $t-th$ column of $H_{ZU}$ with 
$h(t)$.

Using this notation, DeePC and $\gamma$-DDPC can be written exploiting the LQ decomposition of the Hankel data matrix (see e.g., equation (40) in \cite{breschi2022role}) as follows:
\begin{equation}\label{eq:alternative_predictors}
\begin{array}{rcl}
\begin{bmatrix}
z_{ini}\\
u_{f}\\
y_{f}
\end{bmatrix}\!&=&\!\begin{bmatrix}
Z_{P}\\
U_{F}\\
Y_{F}
\end{bmatrix}\alpha\!=\!\begin{bmatrix}
L_{11} & 0 & 0\\
L_{21} & L_{22} & 0\\
L_{31} & L_{32} & L_{33}
\end{bmatrix}
\begin{bmatrix}
Q_{1}\\
Q_{2}\\
Q_{3}
\end{bmatrix}
\alpha\\
& = & \begin{bmatrix}
L_{11} & 0 & 0\\
L_{21} & L_{22} & 0\\
L_{31} & L_{32} & L_{33}
\end{bmatrix}
\begin{bmatrix}
\gamma_{1}\\
\gamma_{2}\\
\gamma_{3}
\end{bmatrix}\!,
\end{array}
\end{equation} 
with $\alpha\in \mathbb{R}^{N}$, $\gamma_{1} \in \mathbb{R}^{\rho(m+p)}$, $\gamma_{2} \in \mathbb{R}^{mT}$ and $\gamma_{3} \in \mathbb{R}^{pT}$. Let us then focus on the case in which $\gamma_{3}$ is bound to be zero and, thus, $\alpha$ is subject to the consistency constraint
\begin{equation}\label{eq:consistency_constraint}
\|(I_N-\Pi)\alpha\|=0,~~\mbox{with }~	\Pi=
H_{ZU}^{\dagger}
H_{ZU}.
\end{equation} 
In this scenario, the predictor used in $\gamma$-DDPC (which corresponds to the one exploited in DeePC) can be rewritten as:
\begin{align}\label{eq:y_hat_gammaDDPC}
\nonumber & \hat y_f(u_f)\!=\!L_{31}\underbrace{L_{11}^{-1}z_{ini}}_{=\gamma_{1}(z_{ini})}+L_{32}\underbrace{L_{22}^{-1}\left(u_{f}\!-\!L_{21}L_{11}^{-1}z_{ini}\right)}_{=\gamma_{2}(u_{f})}\\
& \qquad \quad \!=\! \left(L_{31}\!\!-\!L_{32}L_{22}^{-1}L_{21}\right)\!L_{11}^{-1}z_{ini}\!+\!L_{32}L_{22}^{-1}u_{f}.
\end{align}
As shown in \cite{breschi2022uncertainty}, this (measurable) function of the data $\mathcal{D}$ can be thought of as a (noisy) measurement of the true predictor $\bar y(u_{f})$, as it satisfies a relation of the form:
\begin{equation}\label{eq:noise_pred_measurement}
\hat{y}_{f}(u_{f})=\bar y_f(u_f)+\tilde{y}_{f}(u_{f}).\end{equation} 
The error $\tilde{y}_{f}(u_{f})$, that has been computed in \cite{breschi2023impact}, further satisfies (see also \eqref{eq:predictor_Error}):
\begin{equation}\label{eq:etilde}
	\begin{array}{rcl}
		W\tilde{y}_{f}(u_{f}) &=& \frac{1}{N} E_f H_{ZU}^\top  \hat \Sigma_{ZU}^{-1} \left[\begin{matrix} z_{ini}\\ u_f \end{matrix}\right] \\
		& = &\left(\!\frac{1}{N}\displaystyle{\sum_{i=1}^N e_f(i) h(i)^{\!\top}}\! \right)\!\hat \Sigma_{ZU}^{-1}\!\left[\begin{matrix} z_{ini}\\ u_f \end{matrix}\right]\!, 
	\end{array}
\end{equation}
where 
$\hat \Sigma_{ZU}$ is the sample covariance, namely
$$
\hat \Sigma_{ZU}:= \frac{1}{N} H_{ZU} H_{ZU}^\top = \frac{1}{N}\sum_{i=1}^N h(i)h(i)^\top.
$$

Consider  now the \textquotedblleft true\textquotedblright \ model based predictor in \eqref{eq:predictor3} parametrized as follows
\begin{align}
\underbrace{(I_{pT} -\Phi_{y}^{\gamma})}_{:=W}  \bar{y}_{f}(u_{f})&~\aseq~ \Phi_{P}^{\gamma}z_{ini}+\Phi_{u}^{\gamma}u_{f}  \label{eq:predictor3_PSI}
\end{align}
where the matrices $\Phi_{y}^{\gamma}$, $\Phi_{u}^{\gamma}$ and $\Phi_{P}^{\gamma}$
are defined as
\begin{equation}\label{eq:useful_matrices_gamma}
\begin{array}{c}
		\Phi_{\!u}^{\!\gamma} \!\!=\!\!\! \begin{bmatrix} 
			\!\phi_{1,0}^u &\! \phi_{1,-1}^u  \!&\! \phi_{1,-2}^u \!\!&\! \dots \!&\! \phi_{1,-T+1}^u  \\
			\!\phi_{2,1}^u \!&\! \phi_{2,0}^u \!\!&\!  \phi_{2,-1}^u \!&\! \dots \!& \!\phi_{2,-T+2}^u \\
			\!\phi_{3,2}^u  &\! \phi_{3,1}^u  \!\!&\! \phi_{3,0}^u  \!&\!  \dots \!&\! \phi_{3,-T+3}^u \\
			\!\vdots &\! \vdots  \!&\! \ddots \!&\!  \ddots \!&\! \vdots \\
			\!\phi^u_{T,T\!-\!1}  &\! \phi^u_{T,T\!-\!2}   &\! \dots  \!&\! \phi_{T,1}^u \!&\! \phi_{T,0}^u
		\end{bmatrix}\!\!,\\
		\Phi_{\!y}^{\gamma} \!\!=\!\!\! \begin{bmatrix} 
			\!0 &\! 0  & 0 \!&\! \dots \!&\! 0  \\
			\!\phi_{2,1}^y &\! 0  &  0 \!&\! \dots \!&\! 0 \\
			\!\phi_{3,1}^y  &\! \phi_{3,2}^y  & 0  \!&\!  \dots \!&\! 0 \\
			\!\vdots &\! \vdots  \!&\! \ddots \!&\!  \ddots \!&\! \vdots \\
			\!\phi_{T,T\!-\!1}^y  &\! \phi_{T,T\!-\!2}^y  &\! \dots  \!&\! \phi_{T,1}^y \!&\! 0
		\end{bmatrix}\!\!,\\
		\Phi_{P}^{\gamma} \!=\! \!\!\begin{bmatrix} 
			\phi_{1,\rho} &\phi_{1,\rho-1}  &  \phi_{1,\rho-2} & \dots  & \dots & \phi_{1,1}  \\
			\phi_{2,\rho+1}  & \phi_{2,\rho}  &  \phi_{2,\rho-1} & \dots&\dots &  \phi_{2,2} \\
			\phi_{3,\rho+2} & \phi_{3,\rho+1} &  \phi_{3,\rho}  &  \dots &  \dots&   \phi_{3,3}  \\
			\vdots & \vdots  & \ddots &  \ddots &  \vdots& \vdots   \\
			\phi_{T,\rho+T-1}  & \phi_{T,\rho+T-2}   & \dots  & \phi_{T,\rho+1}  &\dots& \phi_{T,T}
		\end{bmatrix}\!\!. 
\end{array}
\end{equation}
Note that these definitions are similar to those of $\Phi_{y}$, $\Phi_{u}$ and $\Phi_{P}$, with the following differences: $(i)$ each row is parametrized independently, $(ii)$ $\Phi_{u}^{\gamma}$ is not constrained to be lower triangular in the estimation process and 
$(iii)$ all the rows of $\Phi_{P}^{\gamma}$ are fully parametrized and do not have zeros like $\Phi_P$ does. 

By defining the following quantities:
\begin{subequations}\label{eq:auxiliary_def}
\begin{align}
& D_{33}:={\rm diag}[L_{33}],\\
&\hat{W}:=D_{33}L_{33}^{-1},\\
 \nonumber & \tilde{e}_{f}(u_{f}):=W\tilde{y}_{f}(u_{f}) \aseq \hat W\tilde{y}_{f}(u_{f}) \\
&\qquad \qquad \quad  \;= \hat W(\hat{y}_{f}(u_{f}) - \bar y_f(u_{f})),
\end{align}
\end{subequations}
with $L_{33}$ given by \eqref{eq:alternative_predictors} and $\tilde{y}_{f}(u_{f})$ implicitly defined in \eqref{eq:noise_pred_measurement}, we can now establish a relationship between the predictors used in DeePC and $\gamma$-DDPC (see \eqref{eq:alternative_predictors}) and the \textquotedblleft true\textquotedblright \ predictor $\bar y(u_{f})$.
\begin{proposition}\label{prop:equivalnt:statistics:DDPC}
The predictor $\hat{y}_{f}(u_{f})$ satisfies
\begin{equation}\label{eq:WhatY}
\begin{array}{rcl}
\hat W\hat y_f(u_f)& \aseq & W \bar y_f(u_f)+n_f(u_f),
\end{array}
\end{equation}
where \begin{equation}\label{eq:ntilda}
n_f(u_f) := \tilde e_f(u_f)\!-\!\tilde W\hat y_f(u_f),
\end{equation}
while $\hat{W}$ and $\tilde e_f(u_{f})$ are defined as in \eqref{eq:auxiliary_def} and $\tilde W:=W-\hat{W}$.
\end{proposition}
\begin{proof}
See the Appendix.
\end{proof}
Based on these definitions, we can now state the main assumption underlying the subsequent derivation.
\begin{assumption}\label{ass:FP-DDPC}
The matrices \eqref{eq:useful_matrices_gamma} are fully parametrized (i.e., there are no constraints on their structure except for the zeros in $\Phi_y^\gamma$) and the sequences of coefficients $\{\phi_{i,k}^u\}_{k=-T+i,\ldots,\rho+i-1}$ and $\{\phi_{i,k}^y\}_{k=1,\ldots,\rho+i-1}$ are assigned a Gaussian prior with zero mean and  covariance  $K_{\lambda,i}=\lambda K_i$, $K_i=K_i^\top>0$, and are independent over $i$, $\forall i=1,\ldots,T$. 
\end{assumption}

We can then provide the main result of this Section, which is the fundamental starting  
point for deriving the connection between \textbf{Problem~\ref{prob:opt:control_CLOSS}} and DeePC/DDPC.


\begin{theorem}\label{thm:DDPC}
Let Assumption \ref{ass:FP-DDPC}
hold and $L_{t}(u_{f})$ be defined as in \eqref{eq:costdet}. In the non-informative prior limit (i.e., $\lambda \rightarrow \infty$), the following relationship holds:
\begin{equation}\label{eq:cost_equivalence_DDPC}
\mathbb{E}[L_{t}(u_{f})|\hat W\hat y_f(u_f), \hat W] \aseq J_{\gamma}(u_{f})+r_{\gamma}(u_{f}),
\end{equation}
with 
\begin{subequations}\label{eq:L_losses}
\begin{align}
&J_{\gamma}(u_{f})=\|\hat W(y_{r}\!-\!\hat y_f(u_f))\|_{Q}^{2}+\|u_{r}\!-\!u_{f}\|_{R}^{2},\label{eq:DeePC_losses}\\
&r_{\gamma}(u_{f})=\frac{\mathrm{Tr}(Q \Sigma_{\tilde n})}{N},
\label{eq:DeePC_regularizer}
\end{align}
\end{subequations}
where $\Sigma_{\tilde n}$ is the asymptotic variance  of $\sqrt{N}\tilde n_f(u_{f})$ (see equation \eqref{eq:err_deltaW}).

\end{theorem}
\begin{proof}
See the Appendix.
\end{proof}
Note that, in Theorem~\ref{thm:DDPC} we rephrase \textbf{Problem~\ref{prob:opt:control_CLOSS}} by replacing the conditional cost on the full data set ${\mathcal D}$ with a conditional cost that exploits Hankel data matrices in the same way that DeePC/DDPC do (see \eqref{eq:cost_equivalence_DDPC}). Therefore, the equivalences in \eqref{eq:cost_equivalence_DDPC} show that computing the expected cost conditionally on the predictor exploited in DeePC and $\gamma$-DDPC leads to a ``certainty equivalence'' control problem with Tikhonov regularization that is a quadratic function of $u_f$. Note that this last quantity is directly available in closed form thanks to the asymptotic result in Lemma \ref{lem:AsVar} (see the Appendix). 

\subsection{DeePC and $\gamma$-DDPC revisited}
The results presented so far highlight that both DeePC and $\gamma$-DDPC are sub-optimal schemes with respect to the one obtained from  \textbf{Problem~\ref{prob:opt:control_CLOSS}}. Specifically:
\begin{enumerate}
\item as shown by Theorem~\ref{thm:DDPC}, the cost $J_{\gamma}(u_{f})$ in \eqref{eq:DeePC_losses} is retrieved by replacing the (optimal) conditioning on the full data set ${\mathcal D}$ with its (sub-optimal) pre-processed  version, obtained from ``projected'' data;
\item by using \eqref{eq:alternative_predictors} as a predictor, the structure of the  predictor in \eqref{eq:predictor3} is not enforced. In particular, the causality on future control inputs and the (block) Toeplitz structure of $W$, $\Phi_P$ and $\Phi_u$ are not imposed by design.
\item The regularizer $r_\gamma(u_f)$ in Theorem \ref{thm:DDPC} is a weighted $2-$norm, which is not strictly the same  used in DeePC/$\gamma$-DDPC. We show next that, under an extra assumption $r_\gamma(u_f)$ reduces to the $2-$norm regularized used in DeePC/$\gamma$-DDPC.
\end{enumerate}
Specifically, to recover a complete equivalence with $\gamma$-DDPC and DeePC, we introduce the 
following assumption
.
\begin{assumption}{\bf (Output Error noise)}\label{ass:OE}
    The true system is Output-Error, i.e., $\phi_k^{y}=0$ for all $k \in \mathbb{N}$
\end{assumption}


\begin{theorem}\label{thm:specialcase}
Let Assumption \ref{ass:OE} hold, 
let $Q = \hat W^{-\top} Q_o\hat W^{-1}$ in \eqref{eq:L_losses}, $Q_o = I_T \otimes Q_{o,p}$,  $Q_{o,p} = Q_{o,p}^\top \in \R^{p\times p}$ and 
the input 
be independent of past data (i.e., open-loop data collection). Then, under the approximation $y_r\simeq \hat y_f(u_f)$, the regularizer $r_{\gamma}(u_{f})$ in \eqref{eq:DeePC_regularizer} converges  to
\begin{equation}\label{eq:reshaped_reg}
r_{\gamma}(u_{f}) \mathop{\rightarrow}_{N \rightarrow \infty}\frac{\sigma^{2}{\mathrm{Tr}}[Q_o]}{N} \left(\|\gamma_{1}\|^{2}+\|\gamma_{2}\|^{2}\right),
\end{equation}
with $\gamma_{1}$ and $\gamma_{2}$ given by \eqref{eq:alternative_predictors}.
\end{theorem}
\begin{proof}
See the Appendix.
\end{proof}
\begin{remark}[Estimating $\sigma$ from data in $\gamma$-DDPC]
As shown in \cite{breschi2023impact}, the data-driven factor $L_{33}$ in \eqref{eq:alternative_predictors} can be exploited to construct an estimate of $\sigma$ as follows:
\begin{equation}\label{eq:signa_est}
\hat{\sigma}=\frac{\mathrm{Tr}(L_{33})}{pT},
\end{equation} thus making the optimization of \eqref{eq:cost_equivalence_DDPC} (with $J$ and $r$ as in \eqref{eq:L_losses} and \eqref{eq:reshaped_reg})  \emph{independent from any tunable parameter}.  
\end{remark}

Since $\gamma_{1}$ is fixed by enforcing the initial conditions via $z_{ini}=L_{11}\gamma_{1}$, then
\eqref{eq:reshaped_reg} reduces to
\begin{equation}\label{eq:ruf_reg}
r_{\gamma}(u_{f})=\frac{\sigma^{2}{\mathrm{Tr}}[Q_o]}{N}\|\gamma_{2}\|^{2},
\end{equation}
and, thus, only a regularization on $\gamma_{2}$ as in the scheme described in \cite[Section 5.1]{breschi2022uncertainty} is enforced. Under a suitable choice of the regularization scheme and the penalties, the take-home massages from this result are the following.
\begin{enumerate}
\item The conditional cost in \eqref{eq:cost_equivalence_DDPC} comprises a certainty equivalence loss $J_{\gamma}(u_{f})$ and a quadratic regularization $r_{\gamma}(u_{f})$.
\item The $\gamma$-DDPC scheme with regularization on $\gamma_{2}$ presented in \cite[Section 5.1]{breschi2022uncertainty} is equivalent to optimizing the expected cost conditionally on the predictor $\hat y_f(u_f)$ in \eqref{eq:y_hat_gammaDDPC}, but only under an Output Error assumption (see Assumption \ref{ass:OE}). 

\item The regularizer in \eqref{eq:reshaped_reg} is the same as that found in \cite[Eq. (18)]{breschi2023impact} with $\beta_3=\infty$ and $\beta_2 = \frac{\sigma^2{\rm{Tr}[Q_o]}}{N}$, suggesting that 
there is no need to augment the cost with lasso-like regularizations (see \cite[Section IV.C]{dorfler2022bridging}), nor to keep the \textquotedblleft slack variable\textquotedblright \ $\gamma_{3}$ as an optimization variable, which can instead be set to zero. Meanwhile, this result highlights the necessity of imposing the consistency constraint in \eqref{eq:consistency_constraint} when using DeePC, thus, promoting the scheme proposed in \cite[Eq. (18)]{dorfler2022bridging}. 
\end{enumerate}

\section{Numerical examples}\label{sec:examples}
The simulation results\footnote{The MATLAB$^{\text{\textregistered}}$ code developed for this study is available on GitHub at \url{https://github.com/marcofabris92/a-separation-principle-in-d3pc} for public access and replication of the results.} in this Section compare  the closed-loop performance obtained solving \textbf{Problem~\ref{prob:opt:control_CLOSS}} under the assumptions of Theorem~\ref{thm:FCE:NIP} ($Q=q_o \hat W^{-\top}\hat W^{-1}$, as discussed in Remark \ref{rem:Q})) with competitors from the literature. In particular, we include results achieved with the formulations in 
\cite{breschi2023impact}, \cite{breschi2022uncertainty}, and \cite[Section IV.B]{dorfler2022bridging}. 
We remark that computing the cost in Theorem~\ref{thm:FCE:NIP}  does not require any tuning, in contrast with such alternative methods
which instead requires tuning regularization parameters. In our tests, these parameters  are selected performing a grid search and optimizing the closed-loop cost given by and oracle that can perform closed-loop experiments on the true system.\\
Control experiments are performed on a 
benchmark LTI system~\cite{LandauReyKarimi1995}:
$$
\begin{array}{rcl}
x(t+1) & = & Ax(t) + B u(t) + Ke(t) \\
y(t) & = & C x(t) + D u(t) + e(t) 
\end{array}
$$
where
\begin{align*}
& A= \begin{bmatrix}
1.4183  & -1.5894  &  1.3161  & -0.8864 \\
1   &     0     &    0    &     0 \\
0 &   1    &     0    &     0 \\
0     &    0  &  1  &      0
\end{bmatrix}\!,~~B=\begin{bmatrix}
1 \\ 0 \\ 0 \\ 0
\end{bmatrix}, \\	
&C=\begin{bmatrix}
0     &    0  &  0.2826  &  0.5067
\end{bmatrix},~~D=0,\\
&K = \begin{bmatrix}
		0.1784 &
		-0.6523 &
		0.2020 &
		2.2910
	\end{bmatrix}^{\top}.	
\end{align*}
Here $e(t)$ is  a zero mean white noise with variance $\sigma^{2} = 4.81\cdot10^{-3}$ chosen to obtain 
$SNR=20~dB$. 
The training dataset $\mathcal{D}$ of length $N_{data} = 250$ is obtained simulating the system fed with a zero mean, low-pass filtered white noise $u(t)$ (cut-off angular frequency  $1.8~rad/s$.)\\
We perform $100$ Monte Carlo runs over three different setups (specified 
next). 
The length $\hat \rho$ of the predictor  (see equation \eqref{eq:predictor2}) is determined at each run via the AIC criterion \cite{Ljung}, resulting in the 
distribution depicted in \figurename{~\ref{fig:rhoc3SNRlow}}.
The cost parameters ($q_o$, $R$ and $u_r$) are the same utilized in   \cite{breschi2022uncertainty}, i.e.,  {$q_o=1$}, $R=rI_{m\hat{\rho}}$, with $r=5\cdot 10^{-6}$, $u_r=0$. 
The closed-loop experiment is conducted over a time length of $T_{v}=500$ 
and the prediction horizon $T$ (see, e.g., \eqref{eq:future_input_optvar}) is set to $20$. Regularization parameters, where present, may be tuned either \emph{offline} (i.e. testing over a predefined grid by performing, offline, closed-loop experiments on the true plant) or \emph{online}, \emph{without} requiring additional closed-loop experiments.
The three considered scenarios differ by the choice of the reference signal used for tuning regularization parameters and for closed-loop testing: 
	\begin{itemize}
		\item Setup 1: a square wave   reference output 
        $y_{r,1}(t)= \mathrm{square}(2 \pi t/(T_v - 460))$ is used for both tuning and closed-loop testing.
		\item Setup 2: the reference output $y_{r,2}(t)$  in Fig. \ref{fig:BS_sims_new} (black solid line) is used for both tuning and closed-loop testing.
		\item Setup 3:  the tuning of regularization parameters is carried out by tracking the reference output $y_{r,1}(t)$, whereas the control loop is closed by tracking $y_{r,2}(t)$.
\end{itemize}

Closed-loop performance are evaluated using the index \begin{equation}\label{eq:closed-loopperfindex}
	J_{a} = \frac{1}{T_{v}} \sum_{t=0}^{T_{v}-1} \left(\left\|y(t) - y_{r}(t) \right\|^{2} + r\left\|u(t) \right\|^{2}\right),
\end{equation} where the subscript shall encode the DDPC algorithm used namely:  

\begin{itemize}
	\item $a=DeePC$: DeePC scheme with offline tuning \cite[Equation~(25)]{dorfler2022bridging};
	\item $a=\gamma_2$: $\gamma$-DDPC schemes regularizing the squared norm of $\gamma_{2}$ (and $\beta_3 = \infty$) with offline and online tuning (see \cite{breschi2022uncertainty});
	\item $a=\gamma_3$: $\gamma$-DDPC schemes regularizing the squared norm of $\gamma_{3}$ (and $\beta_2 = 0$) with offline and online tuning (see \cite{breschi2022uncertainty});
	\item $a=\gamma_{23}$: $\gamma$-DDPC scheme regularizing both the squared norms of $\gamma_{2}$ and $\gamma_{3}$ with offline tuning (see~\cite{breschi2023impact});
	\item $a=FCE$: scheme based on \textbf{Problem~\ref{prob:opt:control_CLOSS}} under the assumptions of Theorem~\ref{thm:FCE:NIP};	\item $a=thm3$: the scheme with online tuning as from Theorem \ref{thm:specialcase} under the approximation $y_r \simeq \hat y_f(u_f)$.
\end{itemize}
The notation  $\bar{J}_{a}$ and $\hat{J}_{a}$ is used to encode whether offline tuning of regularization parameters was performed ($\bar{J}_{a}$) or not ($\hat{J}_{a}$). 
Offline tuning relies on a 
grid carefully designed to make sure the best performance (i.e., that leading to the minimum closed-loop cost estimate) was obtained. In particular, for the schemes indicated by $a=\gamma_{2}$, $a=\gamma_{3}$ and $a=\gamma_{23}$, the regularization parameters $\beta_2$ and/or $\beta_3$ range in a logarithmically spaced grid $G_{a}^{\beta}$ of $|G_{a}^{\beta}| = 202$ fixed points for $\beta_2 \in [10^{-3},10^{1}] \cup \{0,+\infty\}$ and $\beta_3 \in [10^{-7},10^{-3}] \cup \{0,+\infty\}$. Instead, the regularization parameters $\lambda_{1}$ and $\lambda_{2}$ of DeePC vary over a grid $G_{a}^{\lambda}$ of $11$ fixed points for  $\lambda_{1} \in [10^{-6},10^{-1}] \cup \{0,+\infty \}$ and $\lambda_{2} \in [10^{-7},10^{-2.5}] \cup \{0\}$.\\
To provide a benchmark for the closed-loop performance, we also consider the model predictive control obtained assuming full knowledge of the true model of the system (i.e., solving {\textbf{Problem~\ref{prob:MPC}}) and denote the associated performance index as $J_{MPC}$.

\begin{figure*}[t!]
	\centering
	\subfigure[Estimated truncation rule $\hat{\rho}$]{\includegraphics[height=0.24\textwidth, trim={0.3cm 0cm 1cm 0.2cm},clip]{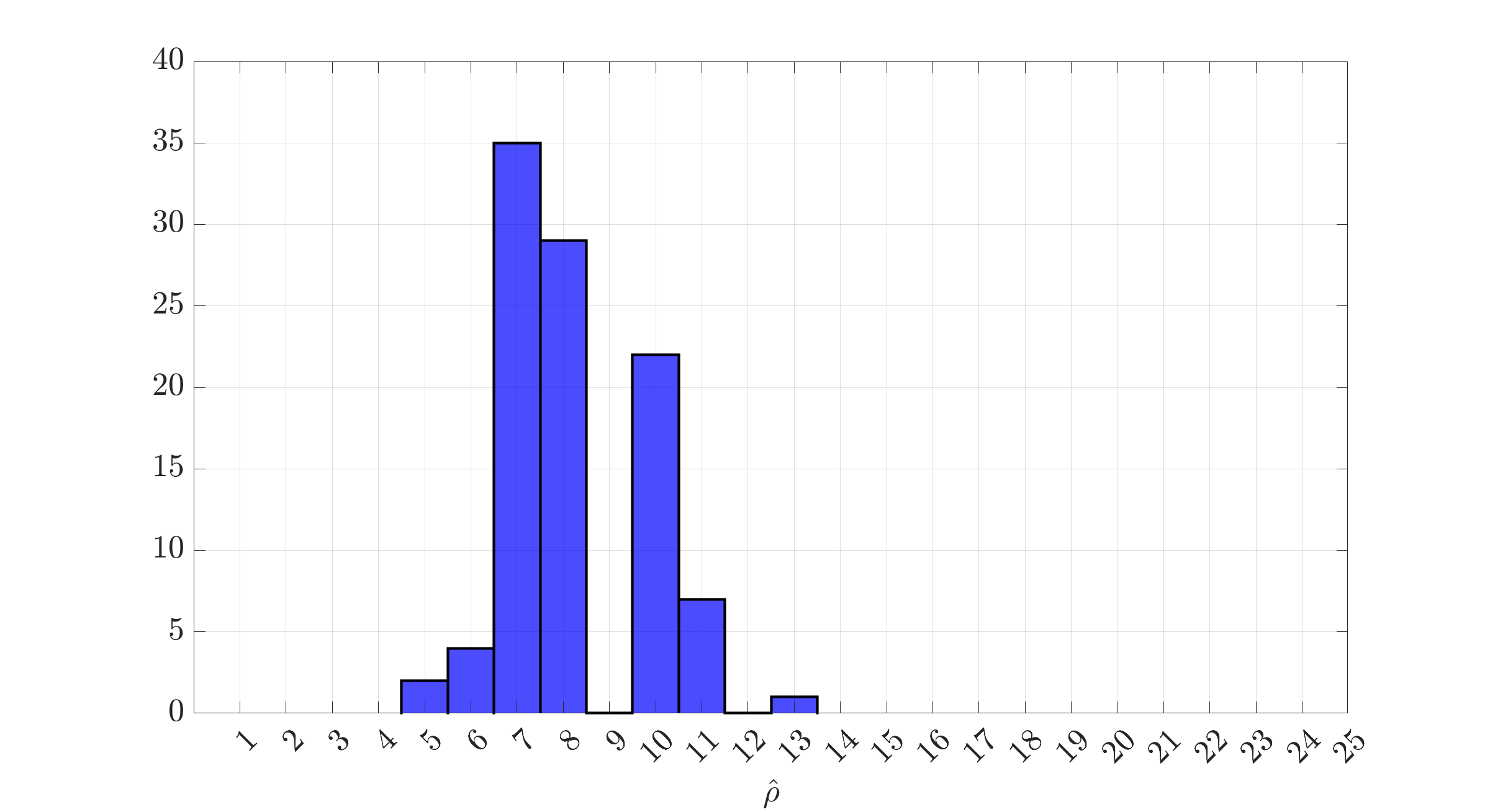}\label{fig:rhoc3SNRlow}}
	\vspace{-.2cm}
	\hspace{5mm}
	\subfigure[Setup 1]{\includegraphics[height=0.24\textwidth, trim={0.3cm 0cm 1cm 0.2cm},clip]{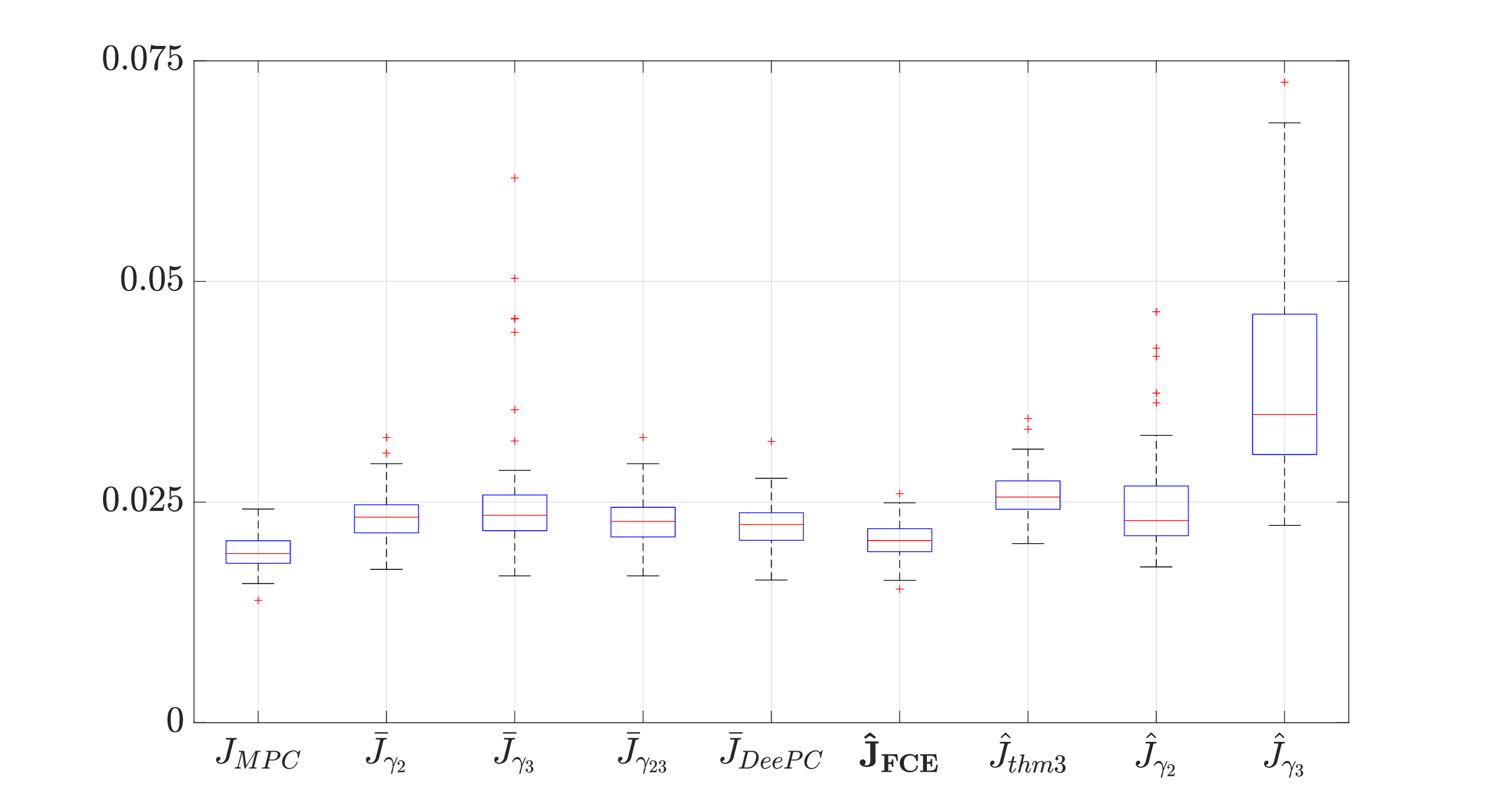}\label{fig:model3_SNR20_A=onda_quadra_regolare_Tv500}}\\
	\subfigure[Setup 2]{\includegraphics[height=0.24\textwidth, trim={0.3cm 0cm 1cm 0cm},clip]{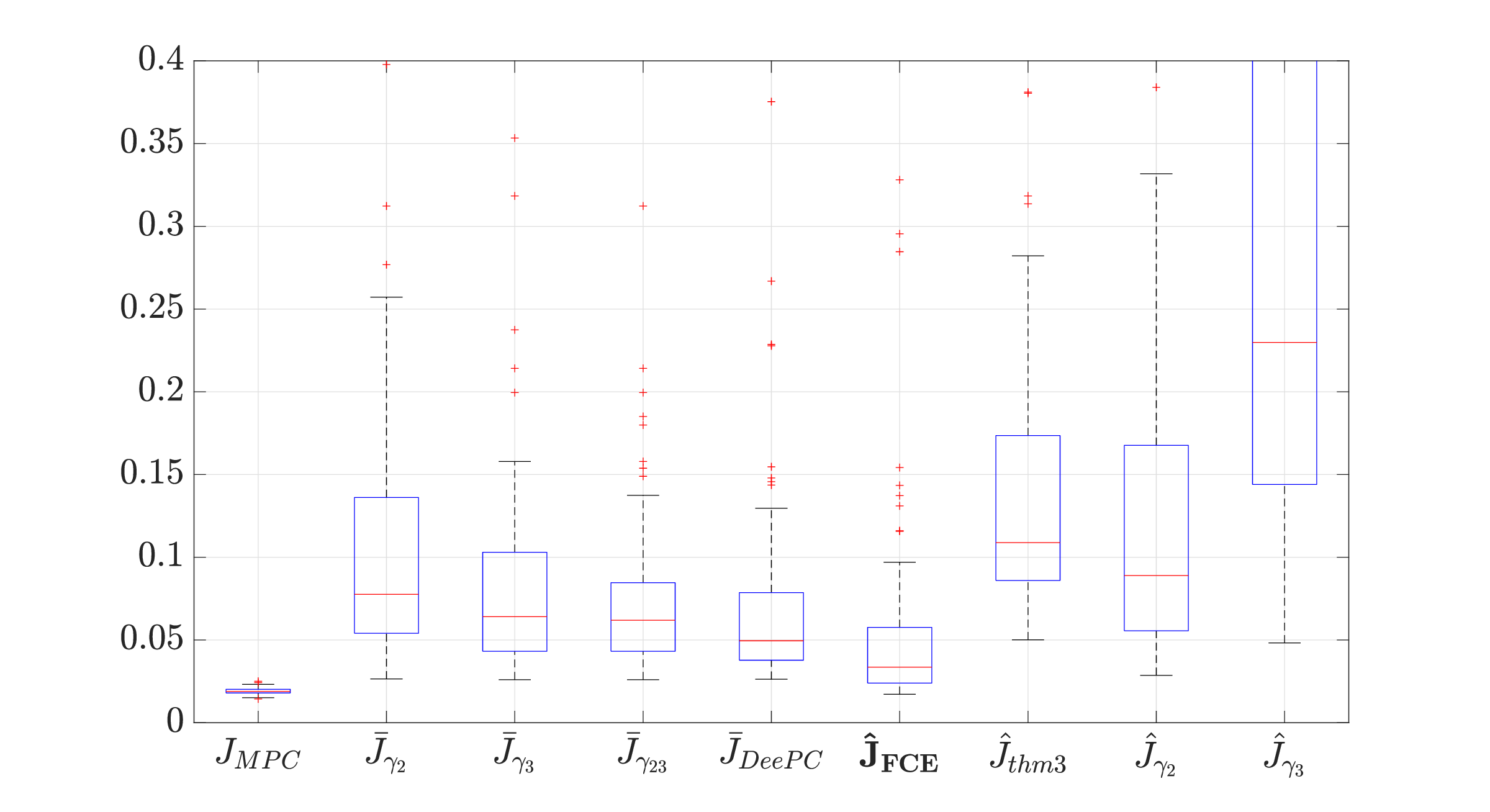}\label{fig:model3_SNR20_B=onda_quadra_irregolare_Tv500}}
	\hspace{5mm}
	\subfigure[Setup 3]{\includegraphics[height=0.24\textwidth, trim={0.3cm 0cm 1cm 0cm},clip]{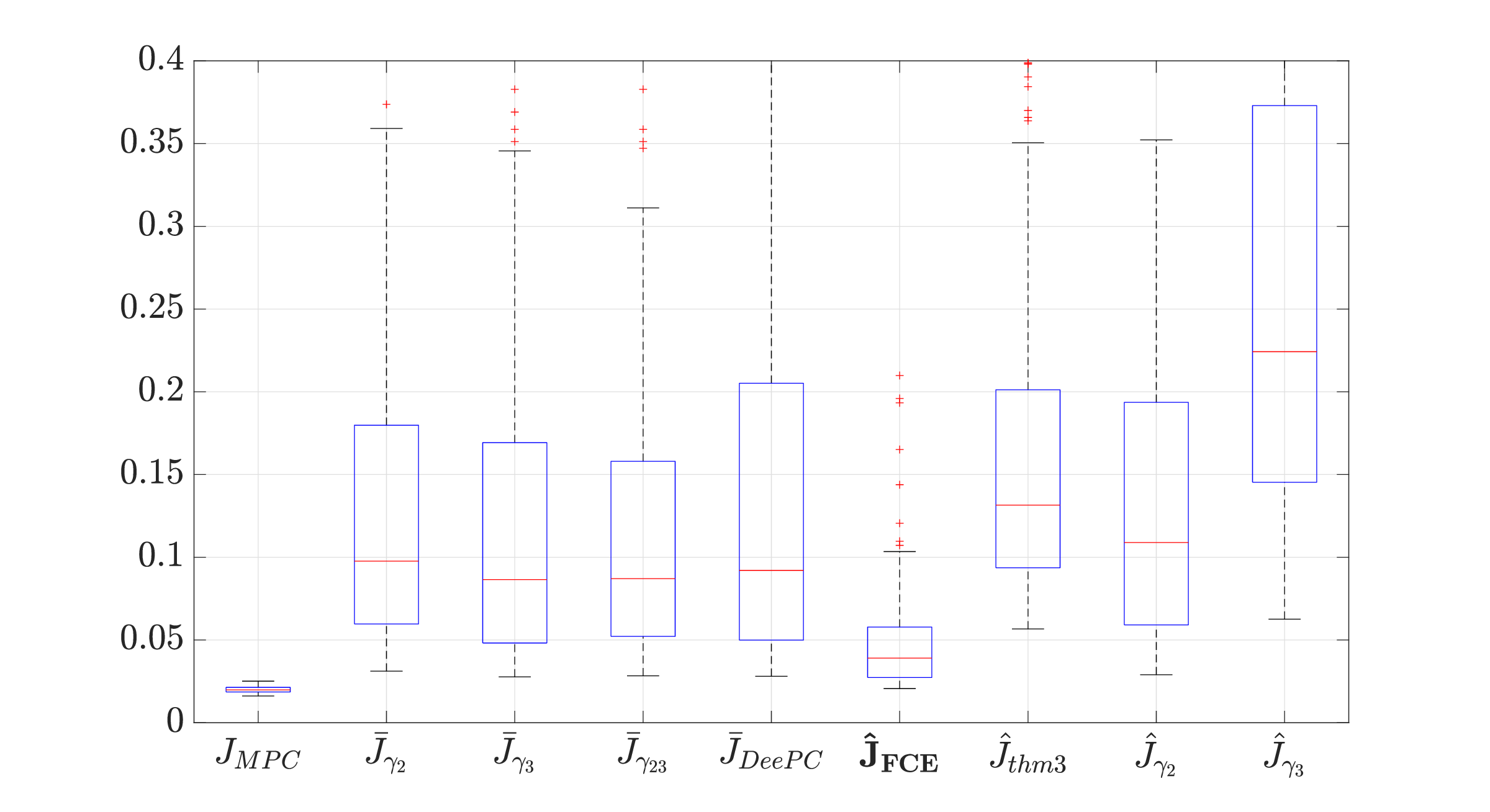}\label{fig:model3_SNR20_tuningA_trackingB}}\vspace{-.2cm}
	\caption{(a): Distributions of the estimated truncation rule $\hat{\rho}$ over $100$ Monte Carlo runs obtained for the Setup 1. (b)-(d): Distributions of the closed-loop performance index (see~\eqref{eq:closed-loopperfindex}) over 100 Monte Carlo runs. Note that the over-lined indexes are used when offline tuning approaches are exploited, whereas indexes with hats refer to the performance attained with online (and, thus, practically feasible) tuning strategies.}
	\label{fig:model_order_distr}
\end{figure*}


\begin{figure*}[t!]
\centering
\subfigure[Setup 3 ($a=DeePC$)]{\includegraphics[height=0.26 \textwidth, trim={0.3cm 0.3cm 1cm 0.2cm},clip]{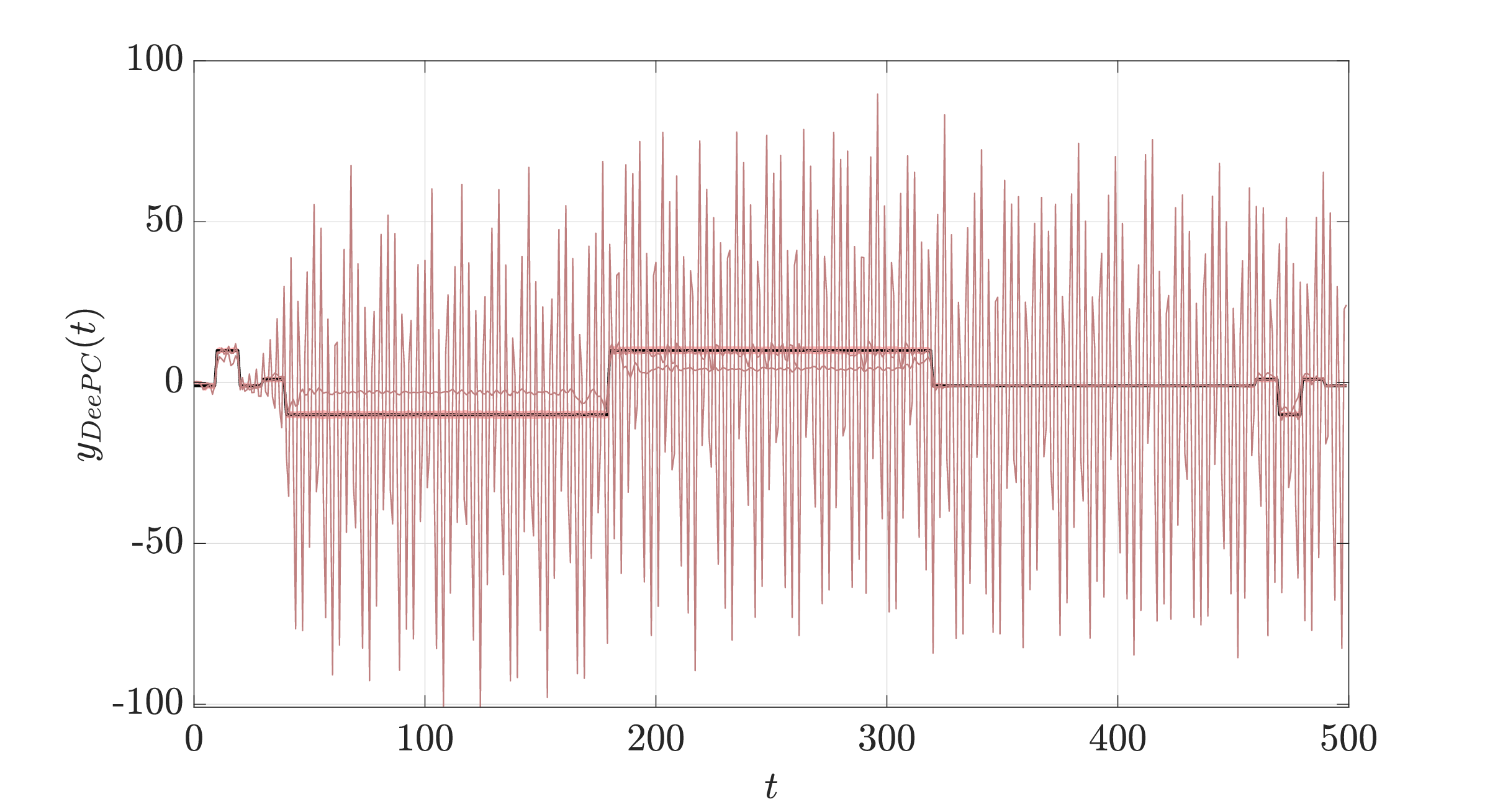}\label{fig:DeePC_50perc_tuningA_trackingB}}
\hspace{5mm}
\subfigure[Setup 3 ($a=FCE$)]{\includegraphics[height=0.26\textwidth, trim={0.3cm 0.3cm 1cm 0.8cm},clip]{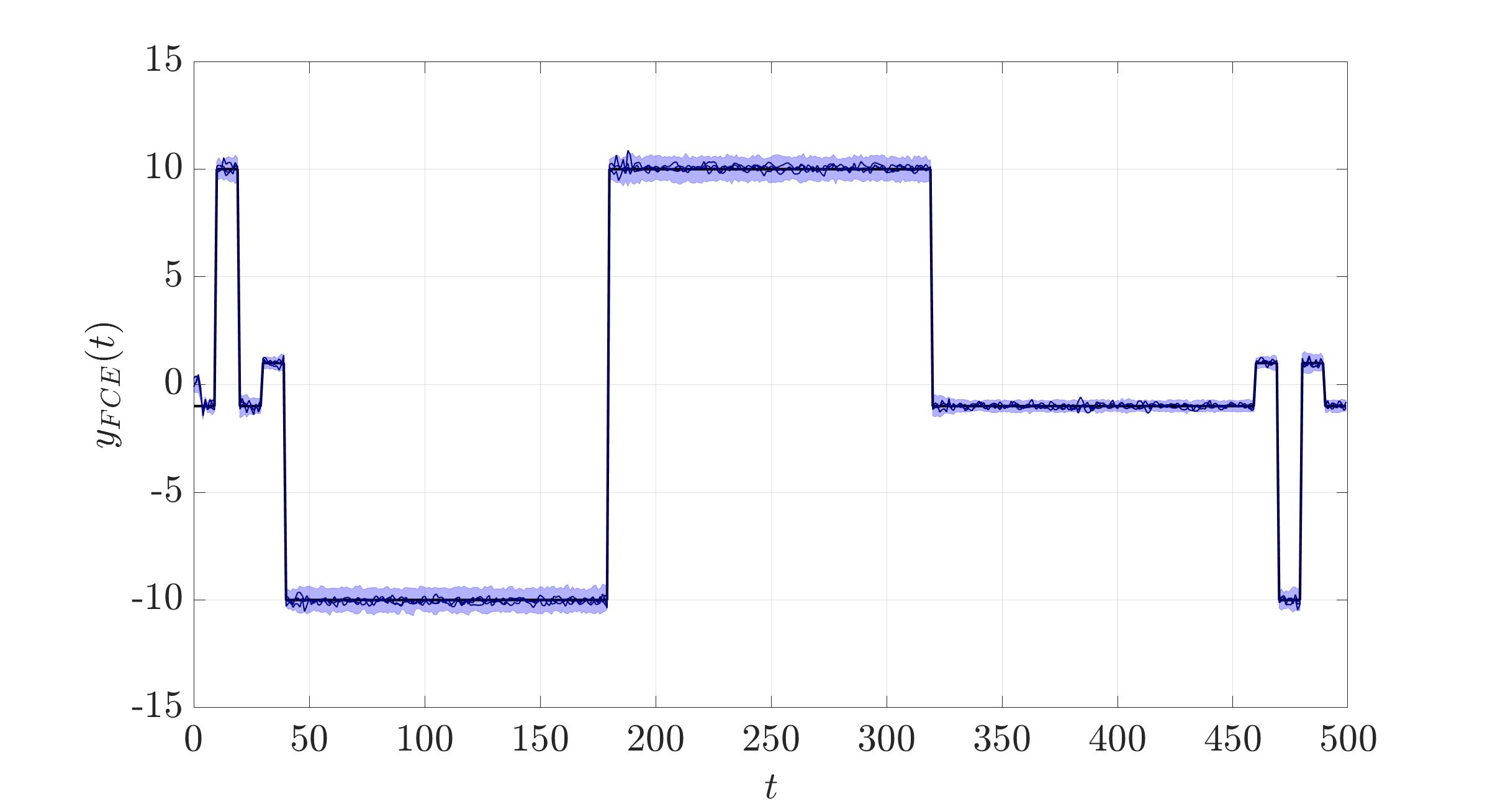}\label{fig:opt_50perc_tuningA_trackingB}}
\vspace{-.2cm}\\
\subfigure[Close-up view of Setup 3 ($a=DeePC$)]{\includegraphics[height=0.26\textwidth, trim={0.3cm 0.3cm 1cm 0.2cm},clip]{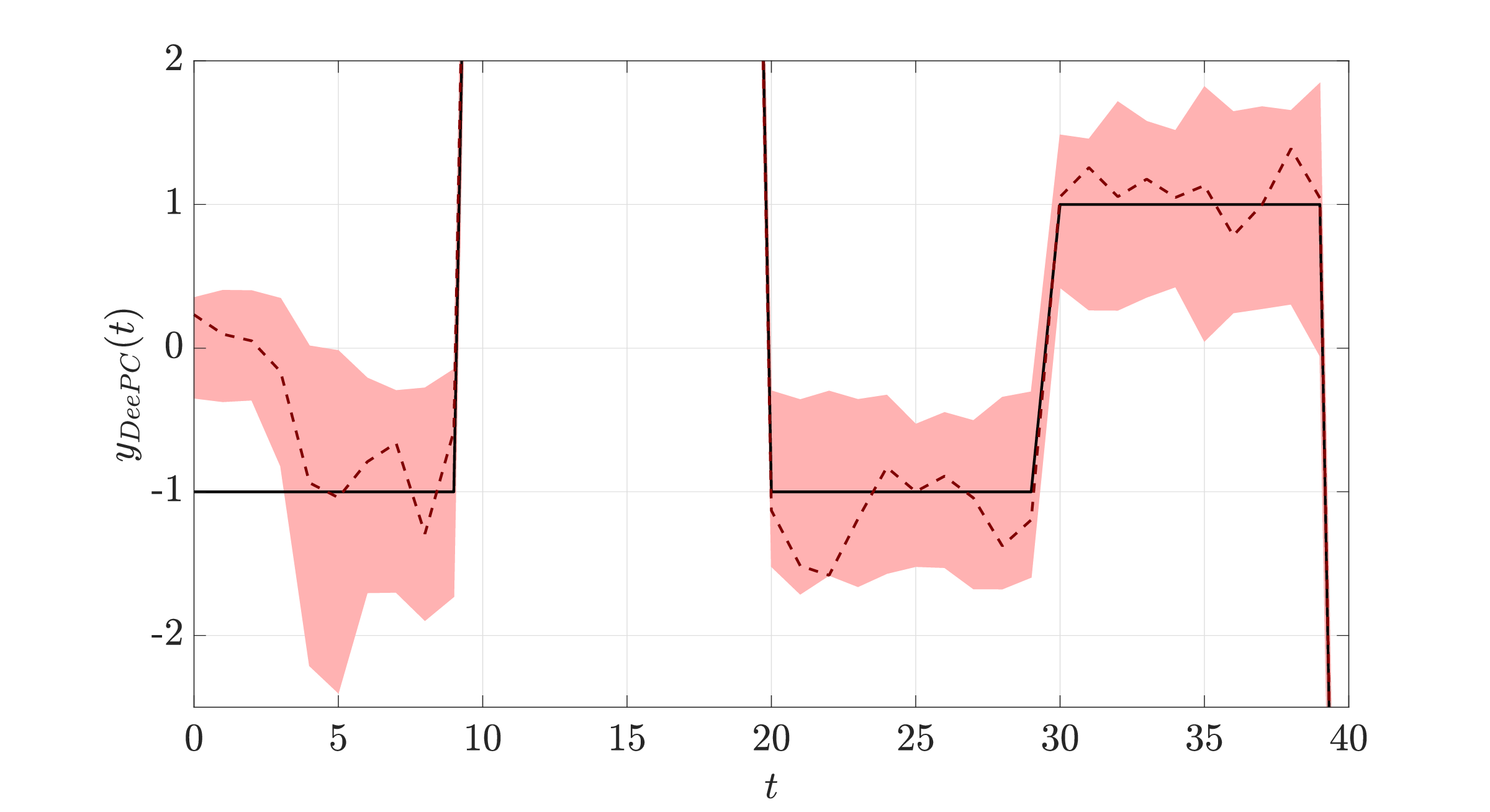}\label{fig:DeePC_50perc_tuningA_trackingB_closeup}}
\hspace{5mm}
\subfigure[Close-up view of Setup 3 ($a=FCE$)]{\includegraphics[height=0.26\textwidth, trim={0.3cm 0.3cm 1cm 0.8cm},clip]{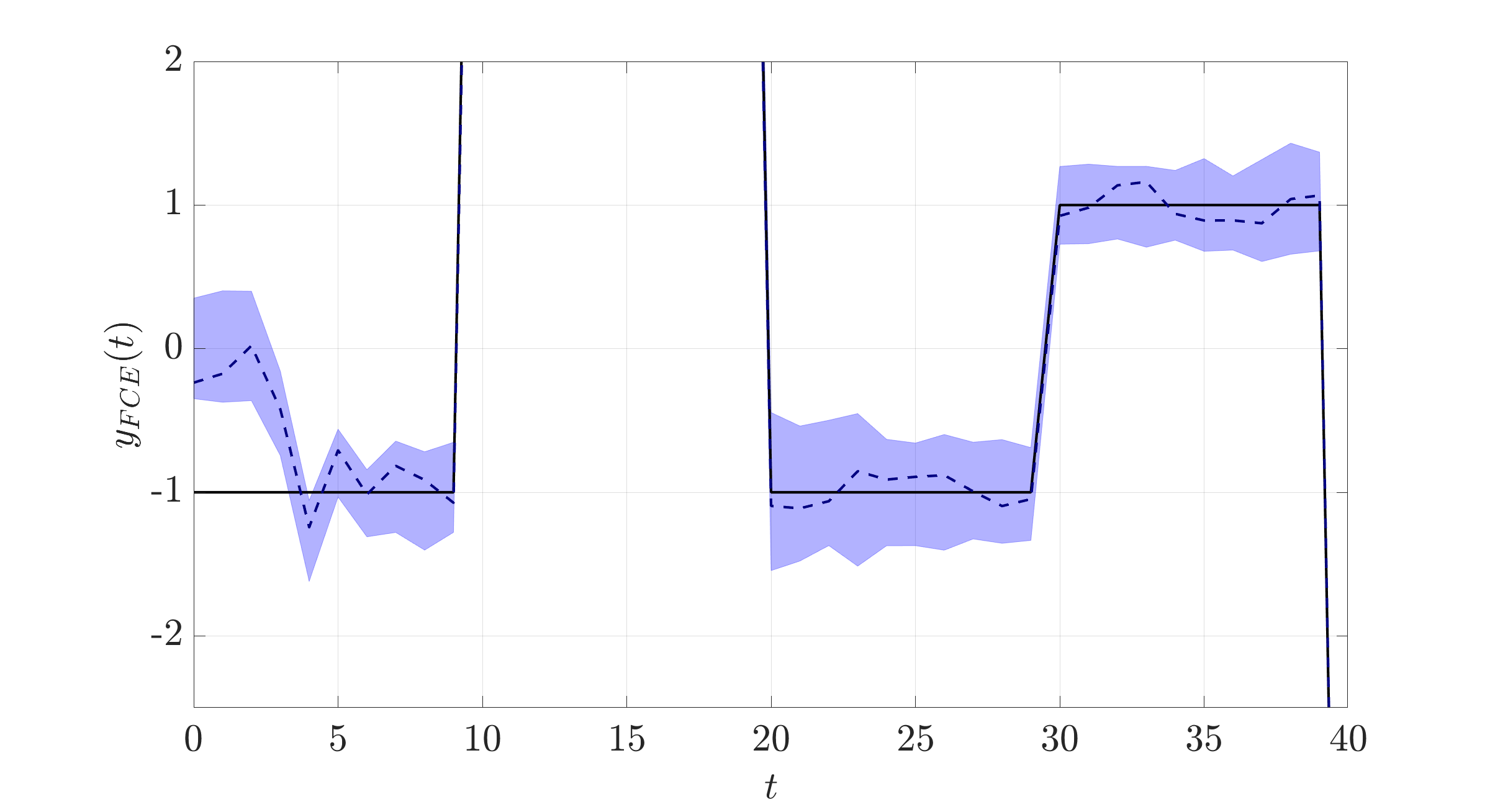}\label{fig:opt_50perc_tuningA_trackingB_closeup}}\vspace{-.2cm}
\caption{Output tracking (solid black line) over  $100$ Monte Carlo runs w.r.t. $a=DeePC$ (red tones) and $a=FCE$ (blue tones). (a)-(b) The 3 worst output trajectories in terms of instantaneous output tracking error adopting the same criterion to draw whiskers in MATLAB$^{\text{\textregistered}}$ boxplots (solid red and green lines). (c)-(d) Median realization in terms of instantaneous output tracking error (dashed red and blue lines); confidence intervals ($\pm 1.96$ std. dev., red and blue shaded areas).}
\label{fig:BS_sims_new}
\end{figure*}


Figs. \ref{fig:model3_SNR20_A=onda_quadra_regolare_Tv500}-\ref{fig:model3_SNR20_tuningA_trackingB} highlights the differences in terms of performance between all the listed methods. 
In particular, we can observe the following.
\begin{itemize}
\item Both offline and online approaches based on the tuning of $\beta_3$ only (i.e., $a=\gamma_{3}$, with $\beta_2=0$) perform significantly worse than the others. 
Even more notably, 
the controlled system becomes practically unstable in $27\%$ 
of the Monte Carlo runs for the Setup 2 (resulting in the index $\hat{J}_{\gamma_{3}}$ being outside  the plotted range in Fig. \ref{fig:model3_SNR20_B=onda_quadra_irregolare_Tv500}
)
.
\item The performance attained with the $\gamma-$DDPC schemes with regularization on $\gamma_{2}$ ($a=\gamma_{2}$) or, jointly, on $\gamma_{2}$ and $\gamma_{3}$ ($a=\gamma_{23}$) 
appear to be essentially equal, as already pointed out in \cite{breschi2023impact}. 
This indicates that, despite the low data regime considered ($N_{data} = 250$), the choice $\beta_3 = +\infty$ (i.e., forcing $\gamma_{3}=0$) is preferable on average.
\item The online strategies $a=\gamma_2,\gamma_3$ perform, as expected, worse than  their offline versions (that use the oracle for parameters tuning)
\item DeePC ($a={DeePC}$) and $\gamma-$DDPC with a regularize on $\gamma_2$ ($a=\gamma_{2}$) perform very similar in all scenarios, suggesting that the $\ell_1$ norm does not make a significant difference in this particular example.
\item The 
tuning-free scheme in \textbf{Problem~\ref{prob:opt:control_CLOSS}} with  regularization as derived in  \eqref{eq:condLossW} ($a=FCE$)
performs on par or better than all the competitors, without requiring any 
tuning. 
\item The (tuning free) version of $\gamma-$DDPC derived in Theorem \ref{thm:specialcase} ($a=thm3$), 
performs worse than  $a=\gamma_2$ 
This is not surprising since we have used the regularizer under the  approximation $y_r \simeq \hat y_f(u_f)$.
\end{itemize}

\begin{table*}[t!]
\caption{Average execution times of the compared predictive schemes ($\pm$ standard deviations) for the Setup 1 and Setup 2.}
\label{tab:centr}
\centering
\scalebox{.85}{\begin{tabular}{|c|c|c|c|c|c|c|c|c|c|}
\hline
a & 
$\gamma_{2}$ (offline) & $\gamma_{3}$ (offline) & $\gamma_{23}$ (offline) & $DeePC$ (offline) & $\mathbf{FCE}$ & $thm3$ & $\gamma_{2}$ (online) & $\gamma_{3}$ (online)\\ 
\hline
training [\si{s}] & 
$0.41  \!\pm\!  0.01$ & $0.41  \!\pm\!  0.01$ & $0.41  \!\pm\!  0.01$ & $0.41 \!\pm\! 0.01$ & $\mathbf{0.42 \!\pm\! 0.01}$ & $0.48 \!\pm\! 0.29$ & $0.48 \!\pm\! 0.01$ &$0.48 \!\pm\! 0.01$ \\
\hline
offline search [\si{s}] & 
$1.15 \!\pm\! 0.00$ & $1.15 \!\pm\! 0.00$ & ${225.67\!\pm\! 0.83}$ & ${7346.5 \!\pm\! 1461.1}$ & $\mathbf{-}$ & $-$ & $-$ & $-$\\
\hline 
optimization [\si{ms}] & 
$56.1 \!\pm\! 1.2$ & $70.0 \!\pm\! 1.1$ & $58.7 \!\pm\! 2.1$ & ${(92.2\!\pm\! 10.7)\!\cdot\! 10^{3}}$ & $\mathbf{64.3 \!\pm\! 0.1}$ & $56.9 \!\pm\! 0.1$ & $3743.0 \!\pm\! 197.5$ & $3099.2 \!\pm\! 208.8$\\
\hline
\end{tabular}}
\end{table*}

\figurename{~\ref{fig:BS_sims_new}} adds to the previous analysis, by illustrating a detailed comparison between the output tracking performance of the proposed approach (with $a=FCE$) and that of DeePC ($a=DeePC$) for Setup 3. These results illustrate that the reference signal plays an important role in tracking, with (i) some outliers suggesting that both near-instability of over-regularization (underfitting) might occur. This behavior does not arise in the proposed strategy (see Figs.~\ref{fig:DeePC_50perc_tuningA_trackingB}-\ref{fig:opt_50perc_tuningA_trackingB}). Moreover, (ii) the output tracking confidence intervals obtained while adopting $a=DeePC$ are approximately twice larger w.r.t. $a=FCE$ (see Figs.~\ref{fig:DeePC_50perc_tuningA_trackingB_closeup}-\ref{fig:opt_50perc_tuningA_trackingB_closeup}). 
Overall, these results indicate that the proposed method outperforms many of 
the recent DDPC approaches.

In light of this discussion, the following key insights can be outlined:
\begin{itemize}
	\item The regularizer \eqref{eq:opt:regularizer} depends on the reference signal and the experimental conditions. This is the main reason why the approach based on \textbf{Problem~\ref{prob:opt:control_CLOSS}} performs better than $\gamma$-DDPC/DeePC even though an oracle is used to tune their regularization parameters. 
	\item The oracle used to select regularization parameters in $\gamma$-DDPC/DeePC  partially compensate for (lack of) dependence of the regularization term on the reference signal and experimental conditions.  Indeed, when tuning is performed on a different reference signal, performance degrades significantly. 
\end{itemize}

Lastly, statistics of CPU times are reported in \tablename{~\ref{tab:centr}}\footnote{Each grid search required for tuning has been parallelized on a 12-core architecture. The DeePC problem has been solved by using the function \emph{fmincon} available in MATLAB$^{\text{\textregistered}}$, imposing a maximum of $100$ function evaluations and a constraint tolerance of $10^{-4}$, exploiting the gradient and Hessian functions of the objective whenever $\lambda_{1} < +\infty$.}. 
We separate the cost for (i) \emph{training} (e.g. computing LQ decompositions/preprocessing/least squares solutions, that needs to be done once for each dataset), (ii) \emph{offline search} (i.e. tuning regularization parameters based on the oracle) and (iii) \emph{optimization} (i.e. the real-time computation of the optimal control, including online tuning when needed, as for $a=\gamma_{2}$ and $a=\gamma_{3}$).

\section{Concluding remarks and future directions}\label{sec:conclusions}
In this work, we have proposed a unified stochastic framework for data-driven predictive control that generalizes existing DDPC methods. The core ingredient of the proposed framework is the Final Control Error, based on which a separation principle for predictive data-driven control is provided. First, a model for the predictor, together with its uncertainty, needs to be computed. Then a regularized quadratic constrained optimization need to be solved. If available, priors on the underlying system can be incorporated. The computational efficiency of the proposed approach, together with the fact that no \textit{ad-hoc} tuning needs to be performed, make the approach practical for real-world applications where noisy data and real-time requirements represent major challenges. \\
Future work will target the incorporation of priors and providing stability and safety guarantees in real-time applications.
\section*{Appendix}
\begin{proofnp}{\bf of Proposition \ref{prop:opt:pred}.}\\
Exploiting linearity of the one step predictor  and using the tower property of conditional expectation, the components $\bar y(t+h|t-1)$ of $\bar y_f(u_f)$, namely
$$
\bar y_f(u_f)= \left[\begin{array}{c} \bar y(t|t-1) \\ \bar y(t+1|t-1) \\ \vdots \\ \bar y(t+T-1|t-1) \end{array} \right], $$
can be recursively obtained from:
$$
\begin{array}{rcl}
\bar y(t+h|t-1) & = & \displaystyle{\mathop{\sum}_{k=1}^{h} \phi_k^y \bar y(t+h-k|t-1)} +  \\ & + &\displaystyle{\mathop{\sum} _{k=h+1}^{\hat \rho} \phi_k^y  y(t+h-k)} +\\
& + & \displaystyle{\mathop{\sum}_{k=1}^{\hat \rho} \phi_k^u  u(t+h-k)}, \end{array}
$$	
for $h\in\{0,\ldots,T-1\}$. The  result in \eqref{eq:predictor3} follows from stacking all these equations 
and exploiting the definitions in \eqref{eq:useful_matrices}. 
Concerning \eqref{eq:predictor_Error}, it suffices to observe that 
$$
y_f \aseq \Phi_P z_{ini} + \Phi_y y_f + \Phi_u u_f +e_f,
$$
from which, recalling the definition of  $W$  in \eqref{eq:model_basedweight},
$$
W y_f \aseq \Phi_P z_{ini} + \Phi_u u_f  +e_f. 
$$
Using \eqref{eq:predictor3},  \eqref{eq:predictor_Error} follows immediately. 
\end{proofnp}

\begin{proofnp}{\bf of Proposition \ref{prop:equivalnt:statistics:DDPC}.}\\
By exploiting \eqref{eq:noise_pred_measurement}, it is easy to show that the following set of equalities hold:
\begin{align*}
\hat{W}\hat y_f(u_f)&=W\hat y_f(u_f)-\tilde W\hat y_f(u_f)\\
& =W\bar y_f(u_f)+W\tilde y_f(u_f) - \tilde W\hat y_f(u_f)\\
& =W\bar y_f(u_f)+\tilde e_f(u_f) -\tilde W\hat y_f(u_f)
\end{align*}
This completes the proof of the equality \eqref{eq:WhatY}. 
\end{proofnp}

The following lemma provides a connection between the LQ decomposition \eqref{eq:alternative_predictors} and a bank of VARX models that will be useful in the subsequent proofs. 

\begin{lemma}\label{lem:LQ_ARX}
Let $\Phi_{i}^{\gamma}$ denote the i-th block rows of the matrix  $[\Phi_{P}^{\gamma}\;\Phi_{u}^{\gamma} \; \Phi_{y}^{\gamma}]$, i.e.,\footnote{Using MATLAB$^{\text{\textregistered}}$-like notation for the block elements defined in \eqref{eq:useful_matrices_gamma}.}:
\begin{equation}\label{eq:Phi_gamma}
\Phi_{i}^{\gamma} : = \left[ \left[\Phi_{P}^{\gamma}\right]_{i,:} \;\; \left[\Phi_{u}^{\gamma}\right]_{i,:} \;\; \left[\Phi_{y}^{\gamma}\right]_{i,1:i-1}\right],
\end{equation}
and consider the VARX models
\begin{equation}\label{eq:bankARX}
Y_{\rho+i} = \Phi_{\gamma,i} \underbrace{\left[\begin{array}{c} Z_P\\ U_F \\Y_{[\rho+1:\rho+i-1]}\end{array}\right]}_{:=Z_{i}^\gamma} + E_{\rho+i},~~~i\!=\!1,\ldots,T.
\end{equation} Let the associated least squares estimators of $\Phi_{i}^{\gamma}$ be:
\begin{equation}\label{eq:LSbankARX}
\hat  \Phi_{i}^{\gamma}:=
Y_{\rho+i}(Z_{i}^\gamma)^\top \left[ Z_{i}^{\gamma}(Z_{i}^\gamma)^\top\right]^{-1}
\end{equation} and build the matrices $\hat \Phi_{P}^{\gamma}$, $\hat\Phi_{u}^{\gamma}$ and $\hat\Phi_{y}^{\gamma}$ defined as in \eqref{eq:useful_matrices_gamma} with the estimated coefficients $ \hat  \Phi_{i}^{\gamma}$. Then, the following equalities hold:
\begin{equation}\label{eq:LinkLQVARX_!}
\begin{array}{rcl}
\hat W  &=&  I_{pT} - \hat \Phi_{y}^{\gamma}, \\
D_{33}Q_{3}& = & Y_F- \hat \Phi_{P}^{\gamma} Z_P -  \hat\Phi_{u}^{\gamma} U_F -  \hat \Phi_{y}^\gamma Y_F, \\ 
L_{31}  &=&  \hat W^{-1} \hat \Phi_{P}^{\gamma}L_{11} + \hat W^{-1}\hat \Phi_{u}^{\gamma}L_{21},\\
L_{32}  &=&  \hat W^{-1} \hat \Phi_{u}^{\gamma}L_{22}, ~~L_{33}  =  \hat W^{-1}D_{33},\end{array}
\end{equation}
and, conversely:
\begin{equation}\label{eq:LinkLQVARX_@}
\begin{array}{rcl}
\hat \Phi_{y}^{\gamma} &=& I_{pT} - \hat W, ~~\hat \Phi_{u}^{\gamma} = \hat W L_{32}L_{22}^{-1},\\
\hat \Phi_{P}^{\gamma} & = & \hat W \left(L_{31}-L_{32}{L_{22}^{-1}L_{21}}\right)L_{11}^{-1},
\end{array}
\end{equation}
with $\hat W:=I_{pT}-\hat\Phi_{y}^{\gamma}$.
\end{lemma}
\begin{proof}
For $i=1,\ldots,T$, let us define the (one-step-ahead) predictors $
\hat Y_{\rho+i}: = \hat \Phi_{i}^{\gamma}$ , which represent the orthogonal projection of the rows of $Y_{\rho+i}$ onto the row space of $Z_{i}^{\gamma}$ (see  \eqref{eq:LSbankARX}). Additionally, define
$$\hat Y_{f,1}:=\begin{bmatrix}\hat Y_{\rho+1}^\top & Y_{\rho+2}^\top &  \cdots &\hat Y_{\rho+T}^\top\end{bmatrix}^\top,$$
that, by exploiting the definition of $\hat \Phi_{P}^{\gamma}$, $\hat\Phi_{u}^{\gamma}$ and $\hat\Phi_{y}^{\gamma}$, 
can be rewritten as:
$$
\hat Y_{f,1} = \hat \Phi_{P}^{\gamma} Z_P + 
\hat\Phi_{u}^{\gamma} U _F + \hat\Phi_{y}^{\gamma}Y_F.
$$
Projecting the last equation on the row space of 
$Z_P$ and $U_F$ and observing that $\hat Y_F := \hat Y_{f,1} \Pi =  Y_F \Pi  $ where $\Pi$ has been defined in \eqref{eq:consistency_constraint}, 
we have that
$$
\hat Y_{F} =  \hat \Phi_{P}^{\gamma} Z_P + 
\hat\Phi_{u}^{\gamma} U _F + \hat\Phi_{y}^{\gamma}\hat Y_{F}.
$$
The projection $Y_F$ can be obtained using the coefficients of the bank of ARX models \eqref{eq:bankARX} as:
\begin{equation}\label{eq:Pred:ARXBANK}
\hat Y_F = \hat W^{-1}\hat \Phi_{P}^{\gamma}  Z_P + \hat W^{-1} \hat\Phi_{u}^{\gamma} U _F.\end{equation}
According to \cite{breschi2022role}, the quantity $\hat Y_F$ can also be computed  using the LQ decomposition as 
\begin{equation}\label{eq:Pred:LQ}
\begin{array}{rcl}
\hat{Y}_F &=& L_{31}Q_1 + L_{32}Q_2 \\
&= & L_{31}L_{11}^{-1}Z_P + L_{32}L_{22}^{-1}(U_F - L_{21}L_{11}^{-1}Z_P)\\
&=& (L_{31}-L_{32}L_{22}^{-1}L_{21})L_{11}^{-1}Z_P + L_{32}L_{22}^{-1}U_F.
\end{array}\end{equation}
Equating the terms of 
\eqref{eq:Pred:ARXBANK} and \eqref{eq:Pred:LQ} the proof straightforwardly follows.
\end{proof}

\begin{proofnp}{\bf of Theorem \ref{thm:DDPC}.}\\
We exploit the decomposition of the second moment in terms of square mean plus variance. Accordingly, by recalling that $\delta_W(u_f)= W(y_r-\bar y_f(u_f))$, the following holds:
$$
\begin{array}{rcl}
\mathbb{E}[L_{t}(u_{f})|{\mathcal D}_\gamma]& = & \| \mathbb{E}[\delta_W(u_f)|{\mathcal D}_\gamma]\|_Q^2 + \|u_{r}-u_{f}\|_{R}^{2} \\
& & + {\mathrm{Tr}}\left[Q {\mathrm{Var}}[\delta_W(u_f)|{\mathcal D}_\gamma] \right],
\end{array}
$$
with ${\mathcal D}_\gamma : = \{\hat W\hat y_f(u_f), \hat W\}$.
We further have that 
$$
\begin{array}{rcl}
\E[ W y_r |{\mathcal D}_\gamma ] & =& \hat W y_r, \\
\E[ W \bar y_f(u_f) |{\mathcal D}_\gamma ] & = & \hat W\hat y_f(u_f), \\
{\mathrm{Var}}[ \delta_W(u_f)|{\mathcal D}_\gamma]  & = & {\mathrm{Var}}[ \tilde{n}_{f}(u_f) |{\mathcal M} ] + o(1/N), \end{array}
$$
where $\tilde{n}_{f}(u_f) $ is defined in \eqref{eq:err_deltaW}.

The first two equations derive from the fact that $\hat W y_r$ and $\hat W\hat y_f(u_f)$ can be seen as unbiased measurements of $Wy_r$ and 
$ W \bar y_f(u_f)$, respectively, leading to the definition of $J_{\gamma}(u_f)$ in \eqref{eq:DeePC_losses}. As it concerns the expression of the conditional variance ${\mathrm{Var}}[\delta_W(u_f)|{\mathcal D}_\gamma]$,  first observe that using \eqref{eq:WhatY} and the definition in \eqref{eq:ntilda} the following holds:
\begin{align}\label{eq:err_deltaW}
\delta_W(u_f) \!-\! \E[ \delta_W(u_f)|{\mathcal D}_\gamma] \nonumber &\!=\!
\tilde W\!(y_r\!-\! \hat y_f(u_f))\! +\! \tilde e_f(u_f)\\ &\!= \! \underbrace{\tilde W y_r + n_f(u_f)}_{:= \tilde n_f(u_f)}
\end{align}
Secondly, notice that the equivalence up to $o(1/N)$ terms derives from the fact that the conditional variance depends upon the estimated predictor (and other sample data covariances). Nonetheless, since these estimators are consistent, they can be replaced by their limits or, equivalently, by the ``true'' ones, as if ${\mathcal M}$ was known. The regularizer in \eqref{eq:DeePC_regularizer} follows by combining these results, concluding the proof. 
\end{proofnp}

The variance term in Theorem \ref{thm:DDPC} can be computed exploiting the following lemma:
\begin{lemma}\label{lem:AsVar}
The asymptotic variance $\Sigma_{\tilde n}$  of the ``noise'' term
$\sqrt{N}\tilde n_f(u_f)$ in \eqref{eq:err_deltaW}
is given by 
\begin{equation}\label{eq:Var:ntilda}
\Sigma_{\tilde n} \!=\! \left[\begin{matrix}\left(\delta(u_f)^{\!\top} \!\otimes\! I_{pT}\right) \!P_W & -I_{pT} \end{matrix}\right] 
\!\Sigma_0\! \left[\begin{matrix}P_W^\top\left(\delta(u_f)^\top \!\otimes\! I_{pT}\right)^{\!\!\top} \\ -I_{pT} \end{matrix}\right]\!, 
\end{equation}
where\footnote{The analytic expression of $\Sigma_0$ can be computed explicitly, but it is rather lengthy and not essential to the purpose of the paper.  Therefore, in the interest of space, it is not reported.}
$$
\Sigma_0 := {\mathrm{AsVar}}\left[\sqrt{N} \left[\begin{matrix}\tilde\theta^\gamma \\ \tilde e_f(u_f) \end{matrix}\right]\right],$$
and
$$
\tilde\theta^\gamma:={\rm vec}\left[\tilde\Phi_{1}^{\gamma},\ldots,\tilde\Phi_{T}^{\gamma}\right],
$$
contains all the parameter estimation errors of the VARX models in \eqref{eq:bankARX}.
\end{lemma}
\begin{proof}
Firstly, let us observe that 
$$
\begin{aligned}
&\tilde{n}_{f}(u_f) :=\tilde W(y_r-\hat y_f(u_f))+  \tilde e_f(u_f)\\
&\quad = \tilde W(y_r- \bar y_f(u_f)) +  \tilde e_f(u_f) + \tilde W(\bar y_f(u_f)-\hat y_f(u_f))\\
&\quad =  \tilde W\underbrace{(y_r- \bar y_f(u_f))}_{\delta(u_f)}+  \tilde e_f(u_f)+ o_P(1/\sqrt{N}),
\end{aligned}
$$
where the last equality is due to the fact that both $\tilde W$ and  $(\bar y_f(u_f)-\hat y_f(u_f))$ are $O_P(1/\sqrt{N})$. The $o_P(1/\sqrt{N})$ terms do not contribute to the asymptotic variance expression and, thus, it can be neglected. Let us introduce a selection  matrix $P_W$ (with entries that are $0$ or $1$) so  that ${\rm vec}[ \tilde W] = P_W \tilde \theta^\gamma$.
Accordingly, it follows that 
$$
\begin{array}{rcl}
\tilde{n}_{f}(u_f) & = & {\rm vec}[\tilde{n}_{f}(u_f)] \\
& = & \left(\delta(u_f)^\top \otimes I\right) P_W \tilde\theta^\gamma + \tilde e_f(u_f),
\end{array}
$$ where the error $\tilde e_f(u_f)= W \tilde{y}_{f}(u_f)$ can be further expressed as in \eqref{eq:etilde}. By exploiting \eqref{eq:LSbankARX}, the error 
${\rm vec}[\tilde \Phi_{i}^\gamma]$ can then be written as: 
\begin{equation}
{\rm vec}[\tilde \Phi_{i}^\gamma] \!=\! \left(I_p \otimes \left[ Z_{i}^{\gamma}(Z_{i}^\gamma)^\top\right]^{-1}\right)  {\rm vec}[ E_{\rho+i}(Z_{i}^\gamma)^\top], 
\end{equation}
ultimately leading to \eqref{eq:Var:ntilda} and, thus, concluding the proof.
\end{proof}

\begin{proofnp}{\bf of Theorem \ref{thm:specialcase}.}\\
The regularizer $r_\gamma(u_f)$ in  \eqref{eq:DeePC_regularizer} can be explicitly obtained from the asymptotic variance $\Sigma_{n}$, whose expression is found in Lemma \ref{lem:AsVar}. In particular, the assumption that $ y_r \simeq \hat y_f(u_f)$ implies that also 
$\delta(u_f) = y_r - \bar y_f(u_f)$ is small, so that $\tilde W \delta$ can be neglected w.r.t. $\tilde e_f(u_f)$. In turn, this result implies that (see \eqref{eq:Var:ntilda}) 
$$
\Sigma_{n} \simeq {\mathrm{AsVar}}\left[\sqrt{N}\tilde e_f(u_f) \right].$$
The asymptotic variance of $\sqrt{N}\tilde e_f(u_f)$ can be computed exploiting  the expression \eqref{eq:etilde}, where $\hat \Sigma_{ZU}$ can be replaced with its asymptotic limit $\Sigma_{ZU}:={\mathbb E}[h(t)h(t)^\top]$ without changing the asymptotic variance (see \cite{vanderVaart}). In particular, by defining
$$
g(i): = h(i)^\top \Sigma_{ZU}^{-1} 
\left[\begin{matrix} z_{ini}\\ u_f \end{matrix}\right],$$
and
$$
\Sigma(i,j): = \E \left[e_f(i)e_f(j)^\top g(i)g(j)\right],
$$
the asymptotic variance of $\sqrt{N}\tilde e_f(u_f)$ can be written as  
$$
{\mathrm{AsVar}}\left[ \sqrt{N}\tilde e_f(u_f) \right] = \mathop{\rm lim}_{N\rightarrow\infty}
\frac{1}{N} \E\left[\sum_{i,j =1}^N \Sigma(i,j)\right].$$
Moreover, by exploiting the martingale difference property of the noises (see Assumption~\ref{ass:Martingale}) and the open loop operation assumption, so that future inputs are independent of past data, we have that
$$
\begin{array}{rcl}
\Sigma(i,j)&: =& \E \left[e_f(i)e_f(j)^\top g(i)g(j)\right] = \sigma^2 J_{i-j} \E \left[ g(i)g(j)\right],
\end{array}
$$
where the matrices $J_{k}$ are defined as follows. Denoting with $[J_{k}]_{i,j}$ the  $(i,j)$-th block ($i,j \in \{1,\ldots,T\}$) of size  $p\times p$ of $J_k $, the following holds:
\begin{equation}\label{eq:Jk}
[J_{k}]_{i,j}=\begin{cases}
I_p, \mbox{ if } j-i = k,\\
0, \mbox{ otherwise.}
\end{cases}
\end{equation}
In particular, note that  $J_k = 0$, $\forall |k|>T$. Therefore the regularization term $r_\gamma(u_f)$ is given by:
\begin{align}
\nonumber r_\gamma(u_f)& \aseq \frac{\mathrm{Tr}(Q \Sigma_{n})}{N} \\
\nonumber &= \displaystyle{\frac{\sigma^2}{N^2}\mathrm{Tr}\left[Q \left(\sum_{\substack{i,j =1\\|i-j|\leq T}}^N J_{i-j} \E [g(i) g(j)] \right)\right]}\\
& = 
\displaystyle{\frac{\sigma^2}{N^2} \sum_{\substack{i,j =1\\|i-j|\leq T}}^N \mathrm{Tr}\left[QJ_{i-j}\right] \E [g(i) g(j)]. }
\end{align}
Under the assumption that the ``true'' underlying system is Output Error, 
we have that 
$$
\hat W \mathop \rightarrow_{N\rightarrow \infty} I_{pT},$$
so that 
$$
\begin{array}{rcl}
\displaystyle{\mathop{\lim}_{N\rightarrow \infty} }\mathrm{Tr}\left[QJ_{i-j}\right]  & = &
\mathrm{Tr}\left[\hat W^{-\top} Q_o\hat W^{-1} J_{i-j}\right]\\
& = & \mathrm{Tr}\left[Q_o J_{i-j}\right] =
\begin{cases} {\mathrm{Tr}}[Q_o] & i=j,\\
0 & i\neq j,
\end{cases}
\end{array}
$$
and, thus, it holds that:
\begin{equation}\label{eq:almostdone}
\displaystyle{\mathop{\lim}_{N\rightarrow \infty} } N r_\gamma(u_f) = \displaystyle{\mathop{\lim}_{N\rightarrow \infty} }\frac{\sigma^{2}{\mathrm{Tr}}[Q_o]}{N} \sum_{i=1}^{N}\E [g(i)^2].
\end{equation}
The expected value $\E [g(i)^2]$ can now be computed as follows: 
$$
\begin{array}{rcl}
\E [g(i)^2] & = & 
\left[\begin{matrix} z^\top _{ini} & u_f^\top \end{matrix}\right]\Sigma_{ZU}^{-1} \underbrace{\E \left[ h(i) h(i)^\top \right]}_{\Sigma_{ZU}}\Sigma_{ZU}^{-1} 
\left[\begin{matrix} z_{ini}\\ u_f \end{matrix}\right] \\
& = & \left[\begin{matrix} z^\top _{ini} & u_f^\top \end{matrix}\right]\Sigma_{ZU}^{-1} 
\left[\begin{matrix} z_{ini}\\ u_f \end{matrix}\right]. 
\end{array}$$
Finally, by recalling that (see \eqref{eq:alternative_predictors}):
\begin{equation*}\label{eq:gamma12}
\begin{bmatrix}
z_{ini}\\
u_{f}
\end{bmatrix}=\underbrace{\begin{bmatrix}
L_{11} & 0\\
L_{21} & L_{22}
\end{bmatrix}}_{=L_{ZU}}\underbrace{\begin{bmatrix}
\gamma_{1}\\
\gamma_{2}
\end{bmatrix}}_{=\gamma_{12}},
\end{equation*}
where
\begin{equation*}
L_{ZU}L_{ZU}^{\top} \underset{N \rightarrow \infty}{\longrightarrow} \Sigma_{ZU},
\end{equation*}
it follows that
\begin{equation}\label{eq:LSigmainv}
L_{ZU}^\top \Sigma_{ZU}^{-1}L_{ZU} \underset{N \rightarrow \infty}{\longrightarrow} I.
\end{equation}
Using \eqref{eq:gamma12} and \eqref{eq:LSigmainv} we finally have that 
$$
\E [g(i)^2]  = \| \gamma_{12}\|^2 = \|\gamma_1\|^2 + \|\gamma_2\|^2
$$
which, inserted in \eqref{eq:almostdone} yields:
\begin{equation}\label{eq:done}
\displaystyle{\mathop{\lim}_{N\rightarrow \infty} } N r_\gamma(u_f) = \sigma^2 {\mathrm{Tr}}[Q_o] \left( \|\gamma_1\|^2 + \|\gamma_2\|^2 \right)
\end{equation}

concluding the proof.
\end{proofnp}

\bibliographystyle{abbrv}
\bibliography{biblio_automatica,biblio_BayesDDPC}

\begin{thebibliography}{10}

\bibitem{akaikefpe}
H.~Akaike.
\newblock Fitting autoregressive models for prediction.
\newblock {\em Annals of the Institute of Statistical Mathematics},
  21:243--247, 1969.

\bibitem{ArtznerCvar1999}
P.~Artzner, D.~F., J.~Eber, and H.~D.
\newblock Coherent measures of risk.
\newblock {\em Mathematical Finance}, 9:203--228, 1999.

\bibitem{Bauer-05}
D.~Bauer.
\newblock Asymptotic properties of subspace estimators.
\newblock {\em Automatica}, 41:359--376, 2005.

\bibitem{Bauer-L-01}
D.~Bauer and L.~Ljung.
\newblock Some facts about the choice of the weighting matrices in {L}arimore
  type of subspace algorithm.
\newblock {\em Automatica}, 38:763--773, 2002.

\bibitem{berberich2020data}
J.~Berberich, J.~K{\"o}hler, M.~M{\"u}ller, and F.~Allg{\"o}wer.
\newblock Data-driven model predictive control with stability and robustness
  guarantees.
\newblock {\em IEEE Transactions on Automatic Control}, 66(4):1702--1717, 2020.

\bibitem{borrelli2017predictive}
F.~Borrelli, A.~Bemporad, and M.~Morari.
\newblock {\em Predictive control for linear and hybrid systems}.
\newblock Cambridge University Press, 2017.

\bibitem{breschi2023impact}
V.~Breschi, A.~Chiuso, M.~Fabris, and S.~Formentin.
\newblock On the impact of regularization in data-driven predictive control.
\newblock In {\em 63rd IEEE Conference on Decision and Control (CDC)}, 2023.

\bibitem{breschi2022role}
V.~Breschi, A.~Chiuso, and S.~Formentin.
\newblock Data-driven predictive control in a stochastic setting: a unified
  framework.
\newblock {\em Automatica}, 152(110961), 2023.

\bibitem{breschi2022uncertainty}
V.~Breschi, M.~Fabris, S.~Formentin, and A.~Chiuso.
\newblock Uncertainty-aware data-driven predictive control in a stochastic
  setting.
\newblock {\em \textnormal{In} 2023 22nd IFAC World Congress}, 2023.

\bibitem{breschi2022design}
V.~Breschi, A.~Sassella, and S.~Formentin.
\newblock On the design of regularized explicit predictive controllers from
  input-output data.
\newblock {\em IEEE Transactions on Automatic Control}, 2022.

\bibitem{CHIUSOvecRegModeling2007}
A.~Chiuso.
\newblock The role of vector autoregressive modeling in predictor-based
  subspace identification.
\newblock {\em Automatica}, 43(6):1034--1048, 2007.

\bibitem{coulson2019data}
J.~Coulson, J.~Lygeros, and F.~D{\"o}rfler.
\newblock Data-enabled predictive control: In the shallows of the {DeePC}.
\newblock In {\em 18th European Control Conference (ECC)}, pages 307--312,
  2019.

\bibitem{Coulson2019}
J.~Coulson, J.~Lygeros, and F.~Dörfler.
\newblock Regularized and distributionally robust data-enabled predictive
  control.
\newblock In {\em 2019 IEEE 58th Conference on Decision and Control (CDC)},
  pages 2696--2701, 2019.

\bibitem{CoulsonLD_DRO2022}
J.~Coulson, J.~Lygeros, and F.~Dörfler.
\newblock Distributionally robust chance constrained data-enabled predictive
  control.
\newblock {\em IEEE Transactions on Automatic Control}, 67(7):3289--3304, 2022.

\bibitem{dorfler2022bridging}
F.~D{\"o}rfler, J.~Coulson, and I.~Markovsky.
\newblock Bridging direct \& indirect data-driven control formulations via
  regularizations and relaxations.
\newblock {\em IEEE Transactions on Automatic Control}, 2022.

\bibitem{DorflerIEEECSM2023}
F.~Dörfler.
\newblock Data-driven control: Part two of two: Hot take: Why not go with
  models?
\newblock {\em IEEE Control Systems Magazine}, 43(6):27--31, 2023.

\bibitem{FAVOREEL1999}
W.~Favoreel, B.~D. Moor, and M.~Gevers.
\newblock Spc: Subspace predictive control.
\newblock {\em IFAC Proceedings Volumes}, 32(2):4004--4009, 1999.
\newblock 14th IFAC World Congress 1999, Beijing, Chia, 5-9 July.

\bibitem{Grimble:92}
M.~Grimble.
\newblock {LOG} optimal control design for uncertain systems.
\newblock {\em IEE Proceedings D (Control Theory and Applications)},
  139:21--30, January 1992.

\bibitem{Hjalmasson-05}
H.~Hjalmarsson.
\newblock From experiment design to closed-loop control.
\newblock {\em Automatica}, 41(3):393--438, 2005.

\bibitem{krishnan2021direct}
V.~Krishnan and F.~Pasqualetti.
\newblock On direct vs indirect data-driven predictive control.
\newblock In {\em 2021 60th IEEE Conference on Decision and Control (CDC)},
  pages 736--741. IEEE, 2021.

\bibitem{LandauReyKarimi1995}
I.~Landau, D.~Rey, A.~Karimi, A.~Voda, and A.~Franco.
\newblock A flexible transmission system as a benchmark for robust digital
  control.
\newblock {\em European Journal of Control}, 1(2):77--96, 1995.

\bibitem{Lazar2022}
M.~Lazar and P.~C.~N. Verheijen.
\newblock Offset–free data–driven predictive control.
\newblock In {\em 2022 IEEE 61st Conference on Decision and Control (CDC)},
  pages 1099--1104, 2022.

\bibitem{Lazar2023}
M.~Lazar and P.~C.~N. Verheijen.
\newblock Generalized data-driven predictive control: Merging subspace and
  {H}ankel predictors.
\newblock {\em Mathematics}, 11(9), 2023.

\bibitem{Ljung}
L.~Ljung.
\newblock {\em System Identification, Theory for the User}.
\newblock Prentice Hall, 1997.

\bibitem{ChiusoMoffatDorfler2023}
K.~Moffat, , F.~D\"{o}rfler, and A.~Chiuso.
\newblock The transient predictor.
\newblock In {\em Proceedings of IEEE CDC 2024 (to appear)}, 2024.

\bibitem{morari1999model}
M.~Morari and J.~H. Lee.
\newblock Model predictive control: past, present and future.
\newblock {\em Computers \& chemical engineering}, 23(4-5):667--682, 1999.

\bibitem{Book_RegID2022}
G.~Pillonetto, T.~Chen, A.~Chiuso, G.~De~Nicolao, and L.~Ljung.
\newblock {\em Regularized System Identification, Learning Dynamic Models from
  Data}.
\newblock Spinger, 2022.

\bibitem{PILLONETTO2011291}
G.~Pillonetto, A.~Chiuso, and G.~{De Nicolao}.
\newblock Prediction error identification of linear systems: A nonparametric
  {G}aussian regression approach.
\newblock {\em Automatica}, 47(2):291--305, 2011.

\bibitem{rakovic2016model}
S.~V. Rakovi{\'c}.
\newblock Model predictive control: classical, robust, and stochastic.
\newblock {\em IEEE Control Systems Magazine}, 36(6):102--105, 2016.

\bibitem{shi2021advanced}
Y.~Shi and K.~Zhang.
\newblock Advanced model predictive control framework for autonomous
  intelligent mechatronic systems: A tutorial overview and perspectives.
\newblock {\em Annual Reviews in Control}, 52:170--196, 2021.

\bibitem{vanderVaart}
A.~van~der Vaart.
\newblock {\em Asymptotic Statistics}.
\newblock Cambridge University Press, 1998.

\bibitem{Vanov-book}
P.~{Van~Overschee} and B.~{De~Moor}.
\newblock {\em Subspace Identification for Linear Systems}.
\newblock Kluwer Academic Publications, 1996.

\bibitem{willems2005note}
J.~Willems, P.~Rapisarda, I.~Markovsky, and B.~{De~Moor}.
\newblock A note on persistency of excitation.
\newblock {\em Systems \& Control Letters}, 54(4):325--329, 2005.

\end{thebibliography}


\newpage

\begin{flushleft}
\begin{minipage}{0.465\textwidth}
    \begin{wrapfigure}{l}{0.22\textwidth}  
    \vspace{-10pt}  
    \includegraphics[width=1.35\linewidth]{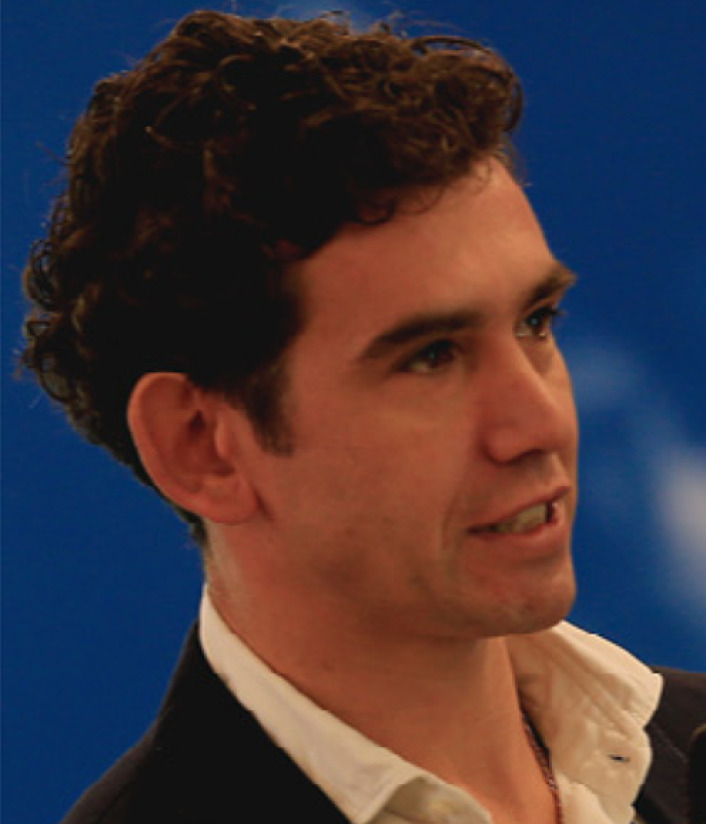}  
    \vspace{-15pt}  
    \end{wrapfigure}
    \textbf{Alessandro Chiuso} received his Master degree (Laurea) in  1996, from the University of Padova and the PhD (Dottorato) in 2000 from the  University of Bologna. He has been long term visitor with several international institutions, among which  Washington University St. Louis, KTH Stockholm, UCLA, ETH Zurich.  He joined the University of Padova  as an Assistant Professor in 2001, Associate Professor in 2006 and then Full Professor since 2017. He currently serves as Editor (System Identification and Filtering) for Automatica. He has served as an Associate Editor for several prestigious journals (among which Automatica, IEEE Transactions on Automatic Control, IEEE Transactions on Control Systems Technology, European Journal of Control) and  has been active in Conference organization (among with General Chair of SYSID 2021, IPC co-chair of SYSID 2024).  He is a Fellow of IEEE (Class 2022). His research interests are mainly at the intersection of Machine Learning, Estimation, Identification and Control.
\end{minipage}
\end{flushleft}

\vspace{-2cm}

\begin{flushleft}
\begin{minipage}{0.465\textwidth}
    \begin{wrapfigure}{l}{0.22\textwidth}  
    \vspace{-10pt}  
    \includegraphics[width=1.35\linewidth]{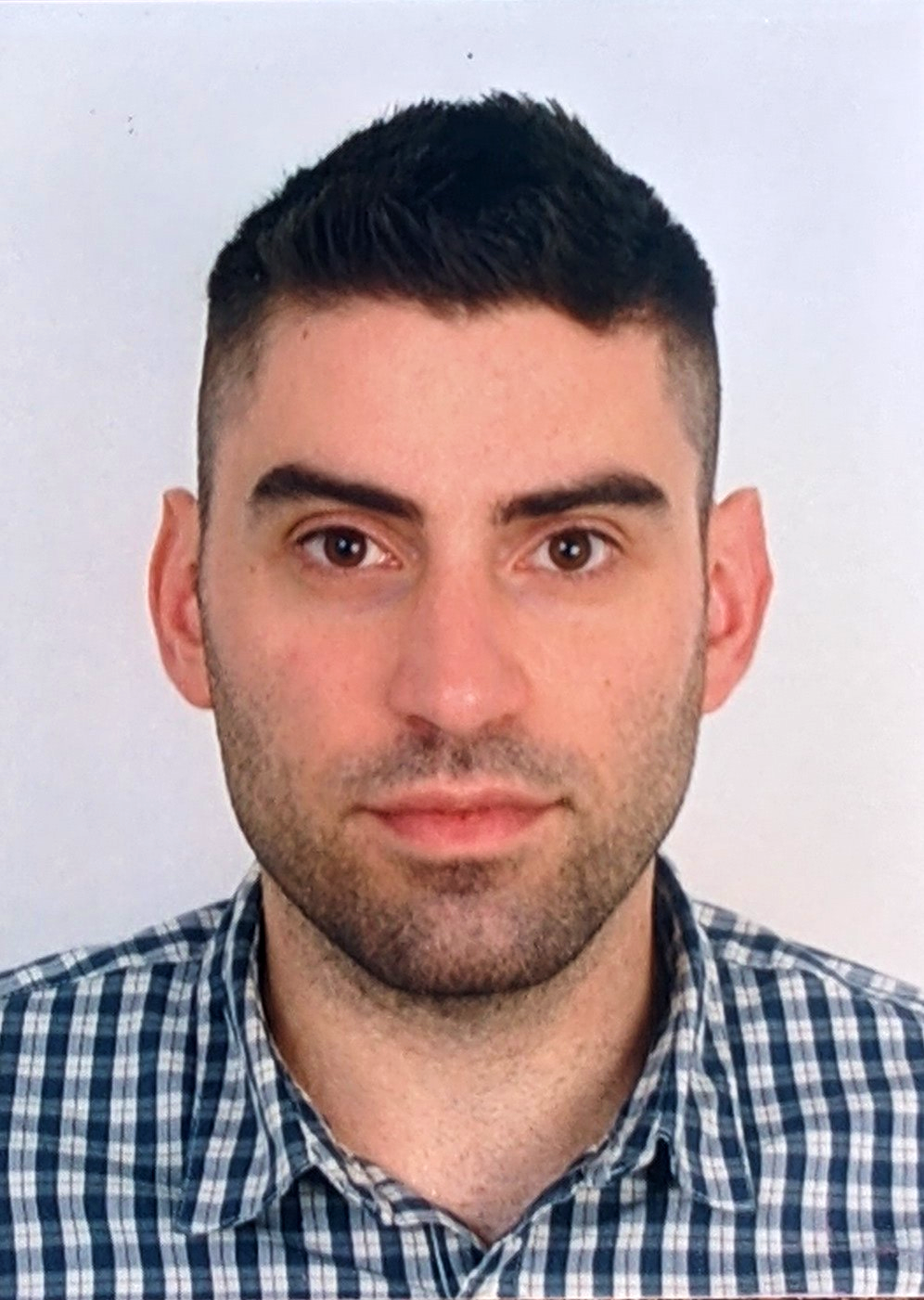}  
    \vspace{-17pt}  
    \end{wrapfigure}
    \textbf{Marco Fabris}
    received the Laurea (M.Sc.) degree (with honors) in Automation Engineering and his PhD both from the University of Padua, in 2016 and 2020, respectively. In 2018, he spent six months with the Electrical, Computer and Energy Engineering Dept. at the University of Colorado Boulder, USA, as a visiting scholar, focusing on distance-based formation tracking. In 2020-2021, he was post-doctoral fellow with the Faculty of Aerospace Engineering at the Technion, Haifa, Israel, directing his attention on the development of secure consensus protocols. From January 2022 to July 2023 he was a post-doctoral fellow with the Dept. of Information Engineering at the University of Padua, where he now holds a position of Research Fellow supported by the Italian National Center for Sustainable Mobility. He is also serving as Associate Editor for the European Control Conference. His main research interests involve multi-agent systems, data-driven predictive control and mobility as a service.
\end{minipage}
\end{flushleft}

\vspace{3cm}

\begin{flushleft}
\begin{minipage}{0.465\textwidth}
    \begin{wrapfigure}{l}{0.22\textwidth}  
    \vspace{-10pt}  
    \includegraphics[width=1.35\linewidth]{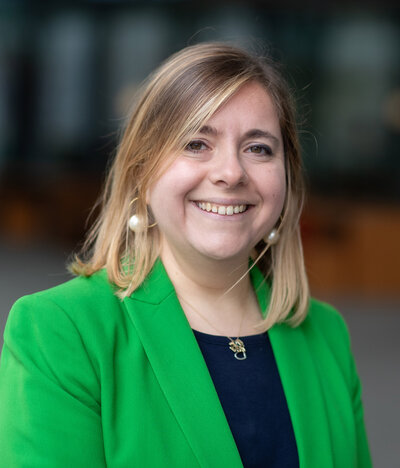}  
    \vspace{-15pt}  
    \end{wrapfigure}
    \textbf{Valentina Breschi} received her B.Sc. in Electronic and Telecommunication Engineering and her M.Sc. in Electrical and Automation Engineering from the University of Florence (Italy) in 2011 and 2014, respectively. She received her Ph.D. in Control Systems from IMT School for Advanced Studies Lucca (Italy) in 2018, being a visiting scholar at the University of Michigan (USA). From 2018 to 2023, she was with Politecnico di Milano (Italy), first as a post-doctoral researcher and then as a junior assistant professor from 2020 to 2023. In 2023, she joined the Department of Electrical Engineering at Eindhoven University of Technology (The Netherlands) as an Assistant Professor. Her main research interests include data-driven control, system identification, collaborative learning, and human-centered policy design, focusing on mobility systems.
\end{minipage}
\end{flushleft}

\vspace{3.24cm}

\begin{flushleft}
\begin{minipage}{0.465\textwidth}
    \begin{wrapfigure}{l}{0.22\textwidth}  
    \vspace{-10pt}  

    \includegraphics[width=1.35\linewidth]{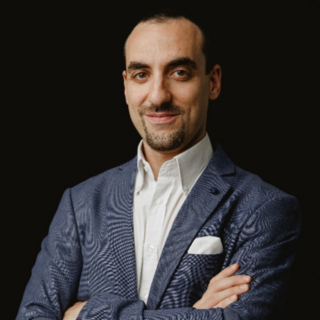}  
    \vspace{-15pt}  
    \end{wrapfigure}
    \textbf{Simone Formentin} was born in Legnano, Italy, in 1984. He received his B.Sc. and M.Sc. degrees cum laude in Automation and Control Engineering from Politecnico di Milano, Italy, in 2006 and 2008, respectively. In 2012, he obtained his Ph.D. degree cum laude in Information Technology within a joint program between Politecnico di Milano and Johannes Kepler University of Linz, Austria. After that, he held two postdoctoral appointments at the Swiss Federal Institute of Technology of Lausanne (EPFL), Switzerland and the University of Bergamo, Italy, respectively. Since 2014, he has been with Politecnico di Milano, first as an assistant professor, then as an associate professor. He is the chair of the IEEE TC on System Identification and Adaptive Control, the social media representative of the IFAC TC on Robust Control and a member of the IFAC TC on Modelling, Identification and Signal Processing. He is an Associate Editor of Automatica and the European Journal of Control. His research interests include system identification and data-driven control with a focus on automotive and financial applications.
\end{minipage}
\end{flushleft}

\end{document}